\documentclass[journal]{IEEEtran}

\usepackage{enumitem}
\usepackage{fancyhdr}
\usepackage{amsmath}
\usepackage{amssymb}
\usepackage{graphicx}
\usepackage{booktabs}
\usepackage{amsfonts}
\usepackage{multirow}
\usepackage{algorithm}
\usepackage{algorithmic}
\usepackage{mathrsfs}
\usepackage{bm}
\usepackage{exscale}
\usepackage{relsize}
\usepackage{setspace}
\usepackage{color}
\usepackage{comment}
\usepackage{ulem}
\usepackage{cite}

\DeclareMathOperator*{\argmax}{arg\,max}
\DeclareMathOperator*{\argminM}{arg\,min}

\ifCLASSINFOpdf
\else
\fi

\begin{document}

\title{Leveraging Channel Knowledge Map for Multi-User Hierarchical Beam Training Under Position Uncertainty}


\author{
		Xu~Shi,~\IEEEmembership{Member,~IEEE},
		Haohan~Wang,
		Yashuai~Cao,~\IEEEmembership{Member,~IEEE},
		Hengyu~Zhang,
		Sufang~Yang,
		Jintao~Wang,~\IEEEmembership{Senior~Member,~IEEE}
		
		\thanks{			
			Xu Shi, Haohan Wang, Hengyu Zhang, and Jintao Wang are with the Department of Electronic Engineering, Tsinghua University, Beijing 100084, China and Beijing National Research Center for Information Science and Technology (BNRist). (e-mail: \{shi-x@, whh24@mails., zhanghen23@mails., wangjintao@\}tsinghua.edu.cn).
			
			Y. Cao is with the School of Intelligence Science and Technology, University of Science and Technology Beijing, Beijing 100083, China (e-mail: caoys@ustb.edu.cn).
			
			S. Yang is with the Future Research Laboratory,
			China Mobile Research Institute, Beijing 100053, China (e-mail: yangsufang@chinamobile.com).
		}
}


\maketitle

\begin{abstract}
Channel knowledge map (CKM) emerges as a promising framework to acquire location-specific channel information without consuming wireless resources, creating new horizons for advanced wireless network design and optimization. 
Despite its potential, the practical application of CKM in beam training faces several challenges. On one hand, the user's precise location is typically unavailable prior to beam training, which limits the utility of CKM since its effectiveness relies heavily on accurate input of position data. On the other hand, the intricate interplay among CKM, real-time observations, and training strategies has not been thoroughly studied, leading to suboptimal performance and difficulties in practical implementation.
In this paper, we present a framework for CKM-aided beam training that addresses these limitations. 
For single-user scenario, we propose a reward-motivated beam-potential hierarchical strategy which integrates partial position information and CKM. This strategy models the user equipment (UE) position uncertainty and formulates the hierarchical searching process as a pruned binary search tree. An optimal hierarchical searching strategy with minimal overhead is derived by evaluating the weights and rewards of potential codewords. Furthermore, a low-complexity two-layer lookahead scheme is designed to balance overhead and computational demands. 
For multi-user scenario, we develop a correlation-driven position-pruning training scheme, where sidelobe gains from inter-user interference are exploited to provide additional side information for overhead reduction, allowing all users to be simultaneously assigned their respective supportive beams.
Simulations validate the superior performances of proposed approaches in advancing 6G beam training.

\end{abstract}

\begin{IEEEkeywords}
Channel knowledge map, beam potential, hierarchical beam training, position uncertainty, training overhead. 
\end{IEEEkeywords}

\IEEEpeerreviewmaketitle

\section{Introduction}

\subsection{Background}
Channel knowledge map (CKM) is poised to deliver an innovative environment-aware communication paradigm for future 6G wireless networks, wielding immense potential to revolutionize various wireless application domains by leveraging location-specific channel information storage \cite{survey1,survey2}. In practical full-spectrum communications including sub-6GHz, millimeter-wave (mmWave) and terahertz (THz) scenarios, CKM acts as a promising enabler to provide additional cost-free spatial/channel prior information and support various critical modules of communication systems such as transceiver deployment, channel state information (CSI) acquisition and beamforming configurations \cite{survey3}. In fact, CKM not only enhances link reliability and reduces latency but also unlocks new possibilities for seamless connectivity in emerging applications like integrated sensing and communication (ISAC), artificial intelligence for radio access network (RAN), immersive extended reality, making it a linchpin for pushing the boundaries of communication technology \cite{survey3,survey4,survey5,beam_training_1}.

However, the practical application of CKM in communication systems remains hindered by significant limitations. Firstly, the efficacy of CKM is heavily constrained by its resolution and input positioning accuracy. Factors such as GPS drift, multipath-induced localization inaccuracy, and dynamic environmental changes (e.g., moving obstacles) degrade the precision of spatial information encoded in the map, leading to wrong decisions and link instability. Secondly, a critical limitation in current literature is the rigid decoupling between CKM and real-time communication observations, where CKM-derived conclusions are usually adopted by treating CKMs as fixed, a priori, and error-free parameters, rather than organic integration that fuses CKMs with real-time observational data. It may cause conflicts between historical information and real-time observations, along with issues of non-real-time performance and poor robustness. Thirdly, existing work on CKM-driven communications has primarily focused on resource scheduling optimization, while its role for promoting CSI acquisition and corresponding strategy design remains significantly underexplored, especially when with consideration of aforementioned limitations. 

\subsection{Related works}

The early study about CKM can be traced back to ray-tracing techniques \cite{CKM_constrct_0} with huge computational complexity. To characterize scattering functions by the presence of multiple objects in real-time, RadioNet \cite{CKM_constrct_1} was firstly proposed with trained deep neural networks in accurate and computationally-efficient manner. After that, approaches like generative adversarial network, diffusion model, and expectation-maximization (EM) algorithm were subsequently proposed in \cite{CKM_constrct_2,CKM_constrct_3,CKM_constrct_4} to enhance construction performance. Then environmental uncertainty and noise were considered based on Bayesian inverse estimation \cite{CKM_constrct_5}, and multi-resolution CKM image \cite{CKM_constrct_6} was further designed by Laplacian pyramid scheme. 
As for the map robustness, interference-cancellation/exploitation  methods \cite{CKM_constrct_ICI_7,CKM_constrct_ICI_8} were studied in the presence of multiple nodes with signal intrusion, and
\cite{CKM_constrct_data_9} investigated the impact of sampling measurements to guarantee the desired level of channel prediction accuracy. Then \cite{CKM_constrct_update_10} studied the self-evolving of CKM via dynamic mode decomposition and \cite{CKM_constrct_MIMO_11} considered the channel fingerprint twin via channel power expectation in scalar form. 
Nonetheless, most works focus on the scalar pathloss map and the extension to multi-degree-of-freedom (DoF) system is quite limited, where \cite{CKM_constrct_MIMO_12} decomposed the channel into several scalar equivalent channels based on orthogonal discrete Fourier transform (DFT) beamforming codebook, thereby enabling the unified construction of high-dimension CKMs. 

As for the applications of CKM, it has been mainly used to optimize communication tasks requiring accurate channel awareness. Initially, CKM enabled environment-aware hybrid beamforming via channel angle/beam index maps \cite{CKM_adopt_beamforming_1}, and was later extended to enhance beamforming in high-mobility networks, RIS scheduling, and UAV scenarios \cite{CKM_adopt_beamforming_2,CKM_adopt_beamforming_3,CKM_adopt_beamforming_4}. 
Notably, all these applications rely on full or partial CSI, with CKM acting as a key carrier for CSI-related insights.
Given CSI’s central role, recent studies have explored how CKM facilitates CSI acquisition. For example, \cite{CKM_adopt_CSI_5} integrated CKM with dynamic sensing to reduce channel estimation uncertainty, while \cite{CKM_adopt_CSI_6} combined CKM and historical CSI for low-pilot channel prediction via joint-orthogonal matching pursuit.
However, traditional channel estimation struggles in large-scale arrays due to exponential pilot overhead. Beam training, by contrast, is more feasible for CSI acquisition here, as it avoids excessive pilots. Yet systematic research on CKM-aided beam training is scarce. Though \cite{CKM_adopt_beam_8} proposed a CKM-enhanced DL-based beam recommendation scheme, it lacks interpretability and strategic control, thus retaining a notable gap from practical deployment.
Besides, CKM is contingent upon user location as input, where its uncertainty and error may undermine the effectiveness, consequently degrading the performance of communication modules.
In essence, CKM is of pivotal significance to beam training. Beyond addressing the pilot overhead challenge in large-scale arrays, it uniquely empowers beam training with CSI-related prior insights, and the balance between these insights and real-time observations might help mitigate CKM’s inherent inaccuracies. Thus it ultimately paving the way for more reliable, robust, and practically deployable beam management.

Specifically with respect to beam training, this constitutes an implicit channel acquisition method as compared to direct channel estimation, while featuring higher reliability and feasibility \cite{beam_training_exhaustive_3,beam_training_2,beam_training_hierarchical_4,beam_training_hierarchical_6,beam_training_hierarchical_near_7,beam_training_hierarchical_5,beam_training_multifinger_9,beam_training_multifinger_8}. In the IEEE 802.11ad and 802.15.3c standards \cite{beam_training_exhaustive_3}, the sector level sweep, beam refinement protocol, and beam tracking mechanisms have been formally established. \cite{beam_training_hierarchical_4} proposed a joint sub-array and de-activation hierarchical codebook based on weighted summation and \cite{beam_training_hierarchical_6,beam_training_hierarchical_5} designed enhanced hierarchical codebook with low sidelobe to confirm training accuracy. Subsequently, multi-finger beam training was devised to enable joint training of multiple users while featuring low overhead \cite{beam_training_hierarchical_5,beam_training_multifinger_9,beam_training_multifinger_8}. The aforementioned beam training methods have ingeniously designed intrinsic codebook patterns and search strategies, yet they do not incorporate auxiliary conditions from external side information such as target and environment characteristics. Furthermore, user motion was incorporated into state transition model and derived Kalman-based fast tracking in \cite{beam_training_Kalman_10}, while sub-6 GHz channel information, historical CSI and real-time measurement was fuzed via neural network to predict the optimal beam \cite{beam_training_deeplearning_11,beam_training_deeplearning_12}. Besides, side information such as user positions \cite{beam_training_side_info_14}, angle parameters from anchor nodes/sensors \cite{beam_training_side_info_13}, as well as scatterers and obstacles in the propagation environment \cite{beam_training_environment_15}, has been exploited to reduce searching range of beam training for overhead reduction and accuracy improvement. Nonetheless, the collaborative coupling within the integrated framework of CKM and beam training has yet to be explored. Existing studies are confined to the impact of either observations or prior information on beam training, while the inherent interplay and  mechanism among the three components, i.e., prior information (from CKM), real-time observations, and training strategy, have not been satisfactorily addressed.

\subsection{Contribution}
\begin{figure}[!t]
	\centering
	\includegraphics[width=0.8\linewidth]{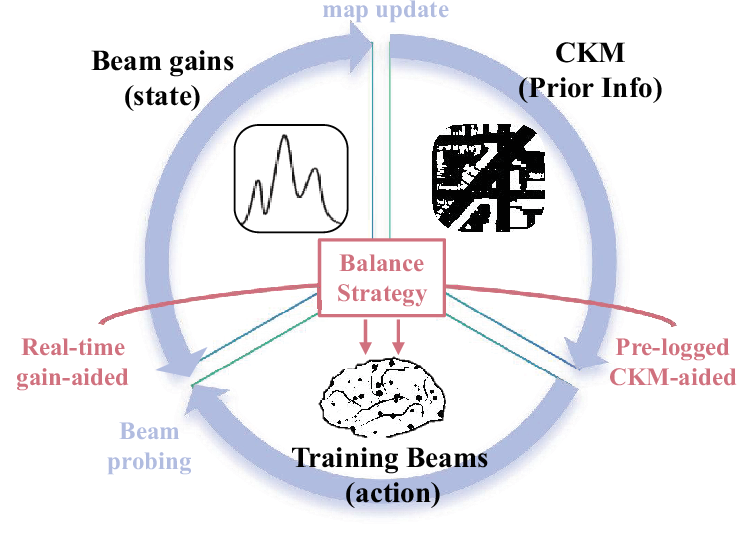}
	\caption{Technical roadmap and logical relationships of CKM-aided hierarchical beam training: CKM uniquely empowers beam training with CSI-related prior insights, while the measured beam gains serve as real-time observation (state) that provide direct guidance for feasible directions. The balance between these insights and real-time observations mitigates CKM's inherent inaccuracy and enhance beam training performance. }
	\label{fig1_question}
\end{figure}

As analyzed above, the main challenges about leveraging CKM for beam training can be summarized as three aspects: 
\textit{(i)} Generally, the user's precise location is not fully known before beam training completed. The reliability of CKM becomes questionable when user position uncertainty exists. For instance, in urban dense areas, blocked line-of-sight (LoS) due to high-rise buildings may restrict GPS accuracy to $50\sim 100$ meters, confining the user to a few city blocks with probabilities weighted by commuting patterns \cite{GPS_error_1};
\textit{(ii)} Careful design is required to strike balance between CKM and observations to determine optimal training strategy, with iterative evolution through recurrent loops, as shown in Fig. \ref{fig1_question};
\textit{(iii)} Factors such as computational complexity, multi-user interference, and training overhead should be carefully considered for the feasibility of CKM-aided beam training. 
In this paper, we put forward a preliminary and constructive solution for these puzzles. The main contributions of this paper are summarized as follows:

\begin{itemize}
	\item Firstly, we propose the concept of beam potential based on CKM and correspondingly provide a reward-motivated beam-potential hierarchical training approach. The partial/low-precision positional knowledge and corresponding CKM are merged as a priori information into beam training procedure to carefully manage the intricate interplay among CKM, observations and strategies. Then we formulate the hierarchical searching process as one complete binary tree, which is pruned using CKM to form a subspace-based incomplete binary tree. We design one layer-coded strategy to select the immediate optimal tree layer and corresponding subspace codewords for low-overhead searching and observation-oriented update.

	\item Secondly, we present a low-complexity two-layer lookahead hierarchical training scheme to further reduce the huge computational and storage complexity of the aforementioned method. During each observation, the number of subsequent search layers and beam codewords are piecewisely determined by considering rewards from only the next two layers of the incomplete binary tree. This scheme is meticulously designed to strike a balance between training overhead and complexity, addressing the bottlenecks that may arise when applying the previous approach in practical scenarios.
	
	\item Thirdly, to address the incompatibility and high overhead of hierarchical beam training in multi-user scenarios, we design one correlation-motivated position-pruning hierarchical training scheme, motivated by correlation between CKM and observations for each user. The sidelobe gains from inter-user interference are additionally exploited to support side information during beam training. The cross-correlation between the prerecorded CKM gain and real-time observed gain is calculated and filtered to prune specific low-probability positional points. This layer-by-layer process gradually shrinks the geographic range until the optimal narrow beam is determined for each UE. 
	
	\item Finally, to demonstrate the efficacy of the proposed CKM-aided beam training algorithms, rigorous simulation verification is conducted. The channel distribution under real-world scenarios is generated using ray tracing method, ensuring the authenticity of the simulation environment. The results clearly indicate that the performance of the proposed algorithms far surpasses that of other existing traditional methods, validating its superiority and practical application potential.
\end{itemize}

\subsection{Organization}
The rest of this paper is organized as follows. In Section \uppercase\expandafter{\romannumeral2}, we introduce the system model and CKM construction. In Section \uppercase\expandafter{\romannumeral3}, we provide the CKM-aided beam training and low-complexity designs for single-user scenario. The multi-user jointly beam training assisted by CKM is proposed in Section \uppercase\expandafter{\romannumeral4}, while the corresponding overhead and complexity analysis is provided. Finally, we show the simulation results in Section \uppercase\expandafter{\romannumeral5} and conclude this paper in Section \uppercase\expandafter{\romannumeral6}.


\section{System Model}

\begin{figure}[!t]
	\centering
	\includegraphics[width=1\linewidth]{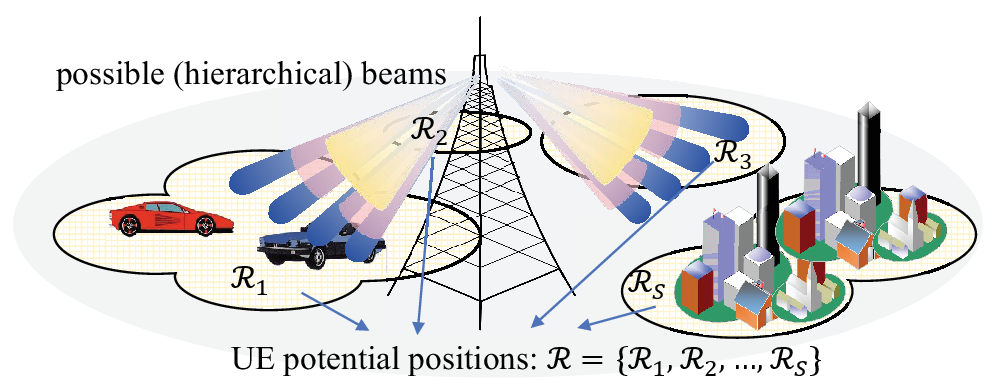}
	\caption{Illustration of the multi-user hierarchical beam training system model.}
	\label{fig2_system_model}
\end{figure}

We consider a general downlink mmWave MIMO communication scenario consisting of one multi-antenna base station (BS) and several single-antenna user equipments (UEs) as shown in Fig. \ref{fig2_system_model}. 
The set of UEs is represented as $\mathcal{K}:\{1,\dots,K \}$. The number of BS antennas is denoted by $N_\text{BS}$, which is arranged to uniform linear array style. We assume the BS is deployed at fixed location $\bm p_0=(x_0,y_0)$. 
Denote $d=\lambda/2$ as the fixed antenna spacing and $\lambda=c/f_c$ represents the wavelength of electromagnetic waves. 
Specifically, the wireless channel from BS to UE $k$ can be formulated as
\begin{equation}
\bm h_k = \sum_{i=1}^{I} g_{k,i} \bm a(\theta_{k,i}),
\end{equation}
where $g_{k,i}$ and $\bm a(\theta_{k,i})=[1,\dots,\text{exp}(-j\pi \theta_{k,i}(N_\text{BS}-1))]$ denote the propagation pathloss and ULA steering vector of the $i$-th path for user $k$, respectively, and $\theta_{k,i}$ represents angle of the departure (AoD) of the corresponding path. The received signal at UE side is written as
\begin{equation}
y_k(\bm f_n) = \bm h_k^H \bm f_{n} s + n_\sigma,
\label{received_signal}
\end{equation}
while $\bm f_n$ represents the $n$-th beamforming codeword, and $n_\sigma\sim \mathcal{CN}(0,\sigma_N^2)$ denotes the additional white Gaussian noise (AWGN) with variance $\sigma_N^2$.

\subsection{Hierarchical beam training}

Before the formal establishment of the downlink connection, beam training is essential for CSI acquisition and UE access. The entire beam training procedure for each UE $k$ can be roughly decomposed into the following steps: 

\begin{enumerate}
	\item BS successively broadcasts potential beams outward at the $0$-th iteration. We define the number of overall search beams as overhead $T$; 
	\item UEs detect the corresponding beam gains, and then feed the related supportive knowledge back to BS; 
	\item BS updates the training strategy and codewords  based on the feedback information and continue the above search. 
\end{enumerate}
The above steps repeat several rounds unitl the optimal beams are found for all users. 
As for the $N_\text{BS}$-antenna ULA, one conventional complete orthogonal beam codebook with cardinal $N_\text{BS}$ can be generated from all columns of DFT matrix $\bm F_{N_\text{BS}}$, and the spatial angles therein are uniformly quantized with spacing $\Delta\theta  = 2/N_\text{BS}$. 

Given that exhaustive searching across the entire beamspace incurs substantial overhead  especially when $N_\text{BS}$ increases sharply, 
hierarchical beam training is widely adopted which mitigates the inefficiencies of exhaustive scanning by progressively narrowing down the search scope through multi-level beam refinement. 
Correspondingly, additional subscripts $l=1,\dots,L$ need to be introduced to denote layer indices in the aforementioned beam searching procedure, i.e., the training codewords $\{\bm f_{l,n}\}$. The number of hierarchical searching layers is formulated as 
\begin{equation}
L = \lceil \log_2 N_\text{BS} \rceil.
\end{equation}
To maintain the consistency and completeness of binary search, the dominant angular range of the $l$-th beam $\bm f_{l,n}$ in the $n$-th layer can be formulated as
\begin{equation}
\text{supp}(\bm f_{l,n}) = \left[ -1 + \frac{n-1}{2^{l-1}}, \, -1 + \frac{n}{2^{l-1}} \right).
\end{equation}

Any two distinct beams within the same layer have disjoint support ranges, i.e.,  $\text{supp}(\bm f_{l,n_1}) \cap \text{supp}(\bm f_{l,n_2}) = \varnothing, \forall n_1\neq n_2$, and this exclusivity guarantees each angular position is covered by at most one beam in each layer, eliminating ambiguity in beam assignment and ensuring the orthogonality of the partitioning structure.

\subsection{BeamCKM concept}

\begin{figure}[!t]
	\centering
	\includegraphics[width=1\linewidth]{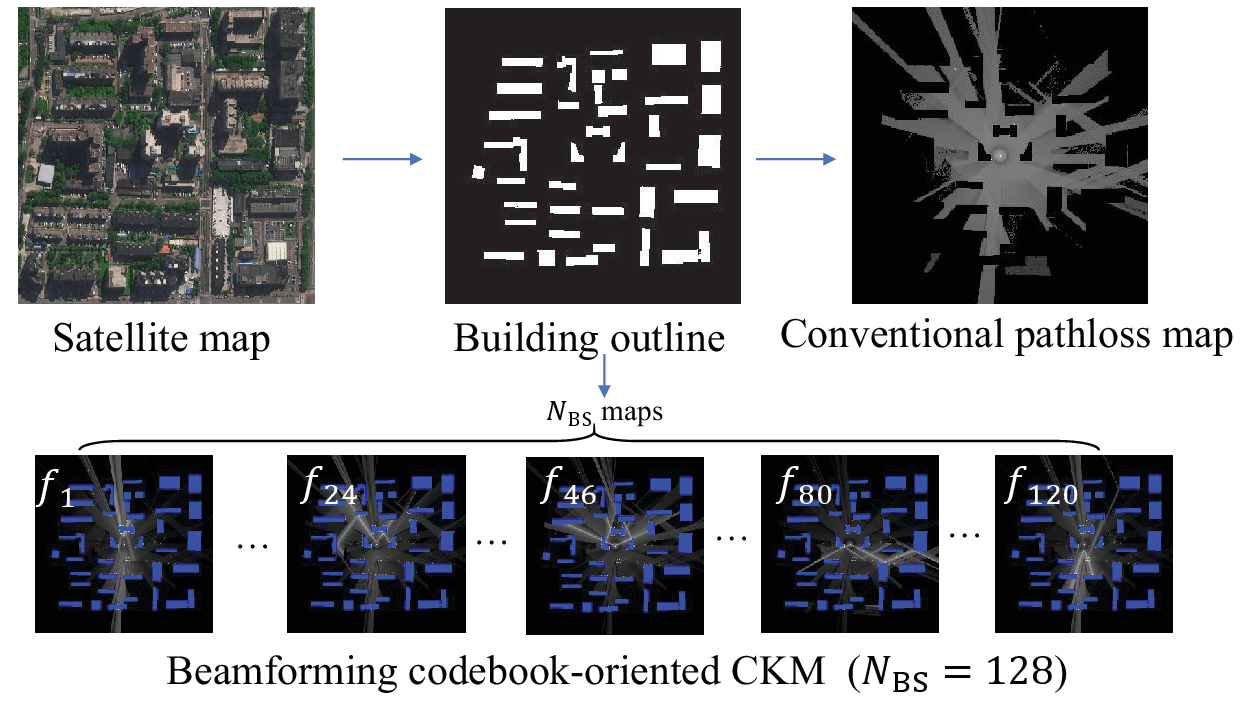}
	\caption{An illustration of beamCKM.}
	\label{fig3_beamforming_CKM}
\end{figure}

CKM offers a novel perspective that provides coarse but significant prior information to assist in the construction of communication links. 
Due to certain objective constraints such as localization bias, user dynamics, and limited accuracy, CKM is promising to preserve some large-scale incoherent parameters instead of coherent phase or accurate CSI. 
However, there exists severe mismatch between single large-scale pathloss map and multiple DoFs in multi-antenna system, and thus we herein adopt the beamforming codebook-oriented CKM, named BeamCKM, to record all propagation gains with each beamforming codeword determined. 

Specifically, we uniformly quantize the complete environment layout with area $X\times Y$ into a number of grid points with horizontal and vertical spacings $\Delta x$ and $\Delta y$, respectively.
This process yields a total of $N_p=\lceil\frac{X}{\Delta x}\rceil\times \lceil\frac{Y}{\Delta y}\rceil$ grid points, collectively forming the overall point set $\mathcal{P}$. Given  beam codeword configuration $\bm f_{l,n}$, we record the equivalent channel gains corresponding to each grid point in set $\mathcal{P}$, thereby constructing the final map. 
In detail, the BeamCKM can be mathematically described as
\begin{equation}
\text{CKM}\big(\bm f_{l,n}\big)\triangleq \mathcal{F}(\mathcal{P}\ |\ \bm f=\bm f_{l,n}).
\end{equation}
For user $k$ with position $\bm p_k$, the recorded equivalent channel gain from CKM can be yielded to
\begin{equation}
G_\text{map}(\bm p_k, \bm f_{l,n})=\mathcal{F}\left(\argminM_{\bm p_{i}\in \mathcal{P}} \ \text{dist}(\bm p_i,\bm p_k) \ \bigg|\ \bm f_{l,n}\right).
\label{gain_map_CKM}
\end{equation}

Although potential supportive beams with strong gain can be directly retrieved via \eqref{gain_map_CKM} from CKM \footnote{It is worth noting that we only focus on the exploitation rather than construction of CKM. Without loss of generality, we assume that the BeamCKM is pre-constructed and stored at BS side before beam training procedure. Details about the construction of BeamCKM are referred to in \cite{CKM_constrct_MIMO_12} for specificity.}, this approach without observation is actually subject to substantial limitations, also with severe position errors. The non-ideality is caused by the following aspects:
\begin{itemize}
	\item \emph{UE positioning uncertainty and mobility:} Before the implementation of beam training and user access, it is unlikely for BS to know the user's exact position due to limited accuracy of localization or integrated sensing and communication (ISAC) technologies, especially considering the UE mobility. It is more reasonable to assume that only partial position information is available for BS. In another word, BS is unable to refine the user's position to a specific grid point but only preliminary determine it within a large-scale region composed of multiple grid points.
	\item \emph{Environmental dynamics and CKM accuracy:} In practice, CKM is pre-stored at BS, whereas the communication environment undergoes real-time changes. We cannot fully rely on the fixed information provided by CKM along the timeline (unless real-time updates to CKM, which constitutes a separate research direction). In fact, a more reasonable approach involves the integration of CKM prior information and real-time observations. Specifically, the prior information derived from CKM serves to provide guidance to searching space and training strategies, while real-time observations are utilized to refine and rectify the deviations inherent in the time-invariant CKM.
\end{itemize}

\subsection{Position uncertainty modeling}

In fact, user localization is often challenged by a combination of technical limitations and environmental factors, leading to a probabilistic representation of discrete zones with nonuniform spatial distribution rather than exact coordinates. 
Specifically, UE position is modeled as a discrete probability distribution over a set of mutually exclusive spatial regions $ \bm{\mathcal{R}}_k = \{ \mathcal{R}_{k,1}, \mathcal{R}_{k,2}, \dots, \mathcal{R}_{k,S} \}$, where $S$ denotes the number of subregions for user $k$ before beam training procedure. Each subregion $\mathcal{R}_{k,s}$ is associated with a prior probability $P_{k,s}$ satisfying $\sum_{s=1}^S P_{k,s} = 1$. 
This distribution reflects the likelihood of the user residing within each region, derived from partial prior information such as mobility patterns, spatial constraints, or historical sensing data.

Each contiguous subregion $\mathcal{R}_{k,s}$ is further discretized into a set of grid points as
\begin{equation}
\mathcal{R}_{k,s}=\{ \bm{p}_{k,s,m} \mid m = 1, 2, \ldots, N_{k,s} \},
\label{contiguous_region}
\end{equation}
where $\bm{p}_{k,s,m}$ denotes uniform-sampled spatial point corresponding to CKM and $N_{k,s}$ denotes the number of grid points inside the subregion $\mathcal{R}_{k,s}$. The grid points within each subregion exhibit distinct spatial characteristics from the BeamCKM, while subregions  are spatially separated from each other (i.e., $\mathcal{R}_{k,s_1} \cap \mathcal{R}_{k,s_2} = \varnothing$ for $s_1 \neq s_2$) due to physical boundaries such as roads, walls, or other natural barriers. 
This two-tier structure combines regional probability distributions and intra-regional grid points, capturing both the ambiguity in coarse localization (at the regional level) and the spatial granularity required for beam training and focusing.

Such a modeling approach is particularly relevant in scenarios where high-precision localization is unavailable, yet contextual constraints limit the user's possible locations to a finite set of discrete regions. For instance, in urban dense areas, blocked line-of-sight (LoS) due to high-rise buildings may restrict GPS accuracy to $50\sim 100$ meters, confining the user to a few city blocks with probabilities weighted by commuting patterns \cite{GPS_error_1}. Similarly, in large indoor spaces like airports, multipath interference in Wi-Fi fingerprinting systems (introducing $10\sim 30$ meter errors) can reduce localization to discrete zones such as departure gates or baggage claims \cite{WIFI_error_2}, with probabilities adjusted based on temporal patterns like boarding schedules. Besides, considering the high-speed mobility scenarios, users may traverse in any direction from a given position, which gives rise to the diversity and uncertainty of predictive positioning \cite{highspeed_error_3}. Even in emergency case (e.g., post-disaster rescue operations with damaged communication infrastructure), weak or fragmented signals may restrict localization to a few residential blocks, with probabilities prioritizing densely populated areas based on prior demographic data \cite{Disaster_error_4}. To summarize, the discrete regional distribution arises from the interplay of limited sensing precision, spatial constraints, and available prior knowledge, making it a practical representation of position uncertainty in real-world communication systems.

\section{Single-user beam training design}
In this section, we focus on the CKM-aided beam training in single-user scenario. We provide the concept of beam potential based on CKM and propose two CKM-assisted hierarchical beam training strategies, where we omit the subscript $k$ to match the single-user case and thus simplify the expression.
The former is oriented toward superior performance via beam-potential reward maximization, while the latter can get a better tradeoff between computational complexity and beam training accuracy via two-layer lookahead. 

\subsection{Reward-motivated beam-potential training}
Based on the nonuniform probability distribution across subregions $\{\mathcal{R}_s\}$, the feasible beamspace can be coarsely reduced via CKM. Thus it is significant to rigorously establish one reward function to evaluate the searching subspace and corresponding optimal layers/codewords during hierarchical training. Then the binary search tree can be pruned with optimal training strategy to reduce overhead. The overall training procedure can be described as the following steps.

\subsubsection{Beam potential and Tree Initialization}
\ 

Firstly, we establish the beam potentials for all codewords via diverse beamforming gains and possible region distributions. When without any observation, the initial potentials of beam codewords should be generated from the prior information inside CKM. After the observations are successively fed back to BS, the binary search tree is updated and corresponding codewords contain dynamic potentials adaptively, where the detailed update will be introduced in the next subsection. Given the UE position uncertainty $\bm{\mathcal{R}}$ and CKM $\mathcal{F}(\mathcal{P}|\bm f)$, we can traverse all possible grid points and obtain the feasible beam distribution $\mathcal{F}(\bm{\mathcal{R}}|\bm f)$. 
We record the grid-wise codeword weight as the product of corresponding equivalent channel gain \eqref{gain_map_CKM} and spatial distribution probability at point $\bm p_{s,m}$, i.e., 
\begin{equation}
\bar{\omega}_{s,m,l,n}\propto p_{s,m}\cdot G_\text{map}(\bm p_{s,m},\bm f_{l,n}).
\end{equation}
This is reasonable that when the probability of a user residing in a specific subregion is high, or when certain beamfocusing gains within one subregion is substantial, larger potential should be assigned to elevate its priority and significance.

Then the bottom-layer beam weights integrated by overall grids can be written as
\begin{equation}
\omega_{L,n} = \sum_{s=1}^S\sum_{m=1}^{N_s}\bar{\omega}_{s,m,L,n} \cdot \mathbb{I}\big(G_\text{map}(\bm p_{s,m},\bm f_{L,n})\geq\Gamma_{s,m} \big),
\label{bottom_layer_weight}
\end{equation}
where $\mathbb{I}(\cdot)$ denotes the indicator function and $\Gamma_{s,m}=\max_n \beta\cdot G_\text{map}(\bm p_{s,m}, \bm f_{L,n})$ and $\beta$ denotes the predetermined threshold to eliminate beams with weak gain, which is consistent with the sparse characteristics of the channel. 

Notice that we only calculate the weights of bottom-layer codewords $\{\bm f_{L,n}\}$ in \eqref{bottom_layer_weight} instead of overall codewords inside $l=1,\dots,L$. This is because the codewords above bottom layer contain a multiplicatively larger beamwidth, resulting in weaker beamforming gain, which means that it is unfair to calculate those weights using the same representation \eqref{bottom_layer_weight}. In fact, since the hierarchical training can be modeled as binary search tree, the weights of top-layer codewords can be calculated as the summation of the weights of leaf nodes, i.e., the bottom-layer codewords, where the recursion formula is written as:
\begin{equation}
\omega_{l,n} = \omega_{l+1,2n-1} + \omega_{l+1,2n},\ \ \ l=1,\dots,L-1.
\label{weighted_codeword}
\end{equation}
For convenience, we collect the hierarchical codewords with strongest several potentials $\omega_{l,n}>0$ and update the candidate codewords as $\{\bm f^+_{l,n}\}$ (also updated iteratively based on observations throughout entire beam searching procedure), where each codeword in $\{\bm f^+_{l,n}\}$ is potential and dominates at certain positions inside CKM.

\begin{figure}[!t]
	\centering
	\includegraphics[width=1\linewidth]{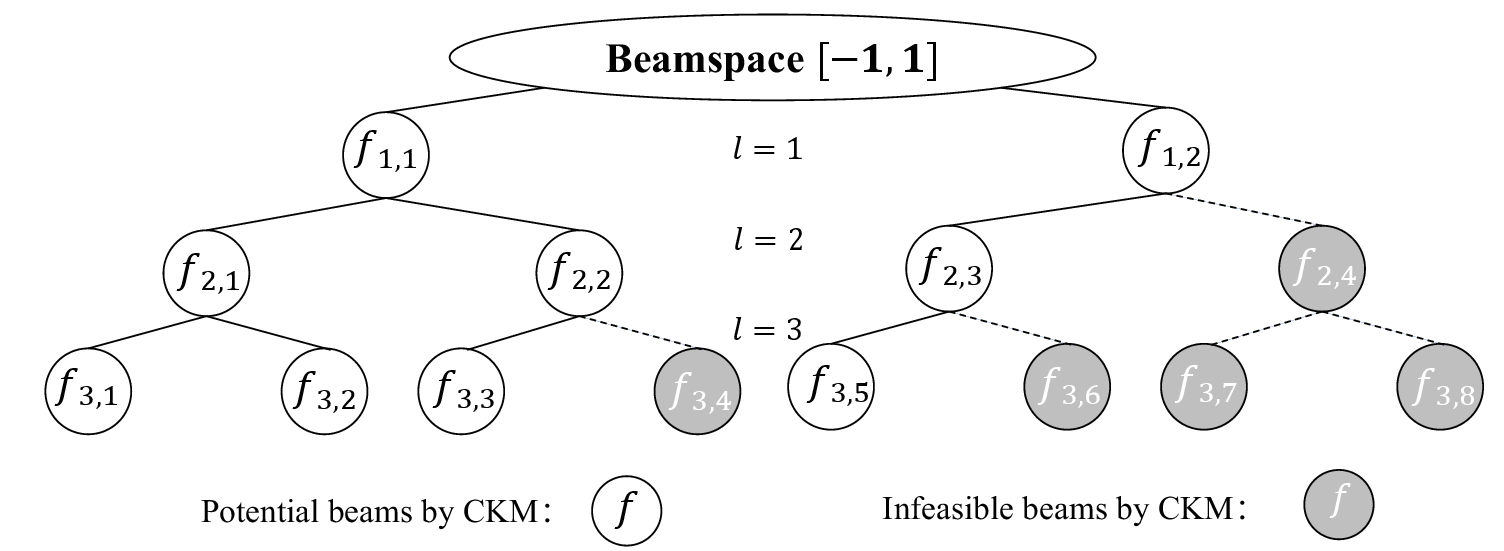}
	\caption{An toy example of pruned search tree for CKM-aided hierarchical beam training.}
	\label{fig4_ToyModel_Tree}
\end{figure}

Given the predetermined potentials of overall codewords from CKM, we can neglect partial codewords with low potential to reduce the searching range. In fact, conventional hierarchical beam training can be equivalently modeled as a complete binary search tree, and thus we can derive one incomplete binary search tree after pruning by CKM. As shown in Fig. \ref{fig4_ToyModel_Tree}, we set configurations $N_\text{BS}=8$ and $L=3$ as a toy model to elaborate on the specific process of CKM-assisted beamspace reduction. Especially when with extremely large array, a large proportion of infeasible beam codewords can be pre-eliminated via CKM, which means the pruned incomplete binary search tree subsequently defines smaller subspaces, thereby ensuring significant overhead reduction. 

\subsubsection{Real-time strategy update}
\

The next question is how to search the incomplete binary tree with smallest overhead. This is highly challenging because optimal search actions vary across different UE locations, which may evolve dynamically with diverse observations of multiple UE positions.
For example, \textit{(i)} when the ideal codeword is $\bm f_{3,5}$ in Fig. \ref{fig4_ToyModel_Tree} from an omniscient perspective, the optimal strategy is to search the only two top-layer codewords $\bm f_{1,1}$ and $\bm f_{1,2}$. Afterwards, we can conclude that $\bm f_{1,2}$ outperforms $\bm f_{1,1}$, and thus the only leaf node, i.e., optimal codeword $\bm f_{3,5}$ is directly determined inside the subspace descending from parent node $\bm f_{1,2}$.  \textit{(ii)} Similarly, when the ideal codeword is $\bm f_{3,3}$ in Fig. \ref{fig4_ToyModel_Tree}, the optimal strategy is to exhaustively search the 2-layer codewords $\bm f_{2,1}$, $\bm f_{2,2}$, and $\bm f_{2,3}$ with overhead consumption $3$, which is smaller than other searching modes like hierarchical searching $\{\bm f_{1,1},\bm f_{1,2},\bm f_{2,1},\bm f_{2,2}\}$ and 3-layer exhaustive searching $\{\bm f_{3,1},\bm f_{3,2},\bm f_{3,3},\bm f_{3,5}\}$. \textit{(iii)} As for ideal codewords $\bm f_{3,1}$ and $\bm f_{3,2}$, 3-layer exhaustive searching is optimal with smallest overhead $4$. {Nonetheless, in practical beam training, the ideal beam cannot be obtained beforehand, because it functions as the ultimate output objective of beam training. To address this, we must take all potential scenarios into consideration and design a unified strategy with maximal rewards to facilitate the observation process.}

To evaluate which layers to start the search from for training overhead reduction, we perform $0$-$1$ encoding for each layer denoted as $\alpha_l$, where $\alpha_l=1$ indicates that feasible codewords inside the $l$-th layer need to be searched and $\alpha_l=0$ represents that codewords inside $l$-th layer are skipped directly. Therefore, we integrate them into activation vectors $\mathcal{A}=\{0,1\}^L$, where the $z$-th activation vector is formulated as
\begin{equation}
\bm \alpha_z=(\alpha_{z,1},\dots,\alpha_{z,l},\dots,\alpha_{z,L})\in \mathcal{A}, \ z=1,\dots,|\mathcal{A}|.
\end{equation}
Since the ultimate output of beam training is to find the optimal narrow beam codeword, the bottom-layer codewords cannot be skipped and thus it always holds that $\alpha_{z,L}=1, \forall z$. The active layer indices corresponding to $\bm \alpha_z$ is written as:
\begin{equation}
	\mathcal{L}_z: \ \{l\in \mathbb{Z}^+\ |\ \alpha_{z,l}=1,1\leq l\leq L\}.
	\label{activation_style}
\end{equation}

Our goal is to find the optimal hierarchical training strategy, i.e., the layer activation and codeword searching policy, thereby decreasing expected overhead for various possible beam directions and observation cases. It is worth noting that strategies and observations are not independent but mutually coupled with each other. Observations require pre-formulated strategies to select potential codewords to search, while strategies need to be adjusted and updated based on real-time observations.

Assume the ideal narrow beam is $\bm f^\text{vir}_\text{opt}=\bm f^+_{L,n}$ from an omniscient perspective, the overall searching codewords via layer-activated strategy $\bm \alpha_z$ through the pruned binary tree can be directly calculated as
\begin{equation}
	\Xi_{z,n}^{0.5}= \sum_{l=1}^L \Xi_{z,l,n}\times\mathbb{I}(\alpha_{z,l}=1),
\end{equation}
where $\Xi_{z,l,n}$ denotes the number of potential codewords in corresponding activated layer $l\in \mathcal{L}_z$ to search for beam  $\bm f^\text{vir}_\text{opt}$ inside set $\{\bm f^+_{L,n}\}$. 
In fact, it can be further divided into two categories. One is the top activated layer $l_\text{min}=\min \mathcal{L}_z$, where all possible codewords that attach to set $\{\bm f^+_{L,n}\}$ should be exhaustively searched to further determine the subtree via observations, and we denote the number as $\Xi^\text{top}_{z,n}$. Correspondingly, we define $\Xi^\text{sub}_{z,l,n}$ as the subsequent child codewords after identifying the optimal parent node in the $(l-1)$-th activated layer, where only codewords inside the pruned subtree should be searched layer by layer $l\in \mathcal{L}_z\backslash l_\text{min}$. Besides, notice that if we determine one codeword at layer $l$ with only one child node in the next activated layer, then it needs not be further searched by directly mapping the codeword into the child node. Therefore, the accurate overhead can be modified as
\begin{equation}
	\Xi_{z,n} = \displaystyle \Xi^\text{top}_{z,n} +  \sum_{l\in \mathcal{L}_z\backslash l_\text{min}} \Xi^\text{sub}_{z,l,n}\times \mathbb{I}( \Xi^\text{sub}_{z,l,n}\geq 2) .
	\label{overhead_calc}
\end{equation}

Considering all possible bottom-layer beams $\{\bm f^+_{L,n}\}$ with non-zero weights $\omega_{L,n}$ in \eqref{bottom_layer_weight}, the merged reward for layer-activation strategy  $\bm a_z$ is derived via weighted summation of overall $\bm f^\text{vir}_\text{opt}=\bm f_{L,n}^+, \forall n$ as
\begin{equation}
	\text{Rew}_{z,0} = - \sum_{n} \omega_{L,n}\cdot \Xi_{z,n},
	\label{reward_1}
\end{equation}
and thus the initial searching layer can be yielded to 
\begin{equation}
l^\text{opt}_0=\min\bigg\{ \argmax_{\mathcal{L}_z}\ \text{Rew}_{z,0}\bigg\}
\label{opt_layer}
\end{equation}

\subsubsection{Beam training and iteration}
\

Following the above initial layer $l^\text{opt}_0$, BS firstly forms directional beams $\{\bm f^+_{l^\text{opt}_0,n}\}$ in several consecutive timeslots. Then UE compares the received signals corresponding to these alternative transmitted beams and sends the codeword index with the largest received power back to BS, denoted as $n_0$. 

In the next steps, consistent to the real-time accurate observation, we refine the searching strategy, where the feasible nodes, weights, strategy space, and reward functions should be correspondingly updated. 
In detail, we revise the temporary root node as the observed one $\bm f^+_{l^\text{opt}_0,n_0}$ and concomitantly retain its child nodes to constrict the search space, where the undetermined hierarchical layers to be searched can be updated to $l=l^\text{opt}_0+1,\dots,L$. The activation vector is modified as 
\begin{equation}
	\bm \alpha_z=(\alpha_{z,l^\text{opt}_0+1},\dots,\alpha_{z,L}).
\end{equation}
In fact, each search round primarily serves to implement a screening-based discrimination that establishes explicit feasibility boundaries to accurately isolate a feasible subset from the extensive candidate beams. Nonetheless, within the resulting feasible beam set,  we rely on in-depth support of CKM $\mathcal{F}(\mathcal{P}|\bm f)$ to achieve quantitative differentiation and dynamic updating of beam potential values (weights) $\omega_{l,n}$, and further provide a precise basis for formulating the subsequent optimal search strategy.

Based on the revised root node $\bm f^+_{l^\text{opt}_0,n_0}$, we adjust the potential of overall codewords in two steps to ensure the execution of subsequent subtree search.  On the one hand, the weights of infeasible codewords (non-child nodes of current root) are set to zero. On the other hand, if there exists a strongest beam direction at any grid point $\bm p_{s,m}$ that conflicts with the observation conclusion $\bm f^+_{l^\text{opt}_0,n_0}$, all point-wise beam weights $\bar{\omega}_{s,m,l,n}$ under the current grid point should be reset to zero to reduce search space.
Then we update the beam potential integrated by overall grids and recalculate the reward function $\text{Rew}_{z,t}$ for each new activation case $\bm \alpha_z$ via \eqref{reward_1}, and yield the subsequent optimal hierarchical layer until the optimal transmitting codeword on the highest layer is determined. To summarize, the detailed expression is provided in Alg.\ref{single_user_algo}.

\begin{algorithm}[htb] 
	\normalem
	\caption{Reward-motivated beam-potential hierarchical training } 
	\label{single_user_algo} 
	\begin{algorithmic}[1] 
		\REQUIRE Hierarchical training codebook $\{\bm f_{l,n}\}$, BeamCKM $\mathcal{F}(\mathcal{P}|\bm f)$, UE positioning prior information including spatial regions $\bm{\mathcal{R}}$ and probability $\{P_{s}\}$, the number of ULA antennas $N_\text{BS}$, predetermined beamforming gain threshold $\beta$, searching round $t=0$. 
		\ENSURE optimal beamforming codeword $\bm f_{L,n_\text{opt}}$

		\leftline{\ \ \ \ \   \textbf{\% Stage 1: CKM-aided initialization}}
		\STATE Calculate the feasible beams $\{\bm f^+_{l,n}\}$ based on CKM $\mathcal{F}(\mathcal{P}|\bm f)$ and positioning range $\bm{\mathcal{R}}$.
		\STATE Calculate the beam potential $\{\omega_{L,n}\}$ based on threshold $\beta$ and CKM gain $G_\text{map}$ via \eqref{bottom_layer_weight}.
		\STATE Prune and update the incomplete binary search tree via weights $\{\omega_{L,n}\}$.
		\WHILE{currently searched layer $l_t<L$}

		\leftline{\ \ \ \ \   \textbf{\% Stage 2: Real-time strategy update}}
		\STATE Capture all activation cases $\{\bm \alpha_z\}$ and corresponding layer index set $\{\mathcal{L}_z\}$ from layer $l_t$ to $L$.
		\FOR{all activation case $z=1,\dots,|\mathcal{A}|$}
		\STATE Traverse all possible beams $\bm f^\text{vir}_\text{opt}=\bm f_{L,n}^+, \forall n$ from a posterior perspective and calculate all accurate overhead $\Xi_{z,n}, \forall n$.
		\STATE Calculate the expected overhead rewards $\mathcal{R}_{z,t}$ via weighted summation in \eqref{reward_1}.
		\ENDFOR
		\STATE Determine the optimal searching layer $l_t=l_t^\text{opt}$ via \eqref{opt_layer}.

		\leftline{\ \ \ \ \   \textbf{\% Stage 3: One-step beam training}}
		\STATE BS searches the directional beams $\{\bm f_{l_t^\text{opt},n}^+\}$.
		\STATE UE sends the codeword index $n_t$ with largest power back to BS.
		\STATE Update the temporary beam $\bm f_{l_t^\text{opt},n_t}^+$ and update search tree based on its child nodes.
		\STATE Searching time $t=t+1$.
		\ENDWHILE
	\end{algorithmic}
\end{algorithm}

\subsection{Low-complexity two-layer lookahead scheme}

In fact, Alg.\ref{single_user_algo} entails a considerably high computational complexity and storage consumption. This is attributed to the global-optimal necessity of traversing all potential possibilities of layer activation $\{\mathcal{L}_z\}$ as well as all potential bottom-layer beams $\{\bm f_{L,n}^+\}$, thereby computing and storing a substantial amount of overhead values $\Xi_{z,n}$. Concurrently, during the hierarchical beam training process, these parameters need to be iteratively updated based on UE's observation and feedback, which will further impinge upon energy efficiency and real-time responsiveness at BS side. Therefore, we provide the low-complexity design for CKM-aided hierarchical beam training, where we focus on formulating strategies for the subsequent two layers of the search tree, based on CKM and observations.

In contrast to the aforementioned global-optimal enumeration method, we neglect the topological structure of the whole binary tree but focus solely on nodes inside the upcoming two layers.
As is well known, each node's subtree constitutes an independent substructure, and the relationship between a parent node and its children is transmitted solely through direct connections. This characteristic allows local information (e.g., the structure of the next two layers) to potentially provide sufficient basis for decision-making. Additionally, the gains from decision-making show a diminishing trend as the number of lookahead steps increases. Therefore, limiting the number of lookahead steps helps balance complexity and performance, thus demonstrating its rationality and effectiveness in practical beam training.
\begin{figure}[!t]
	\centering
	\includegraphics[width=1\linewidth]{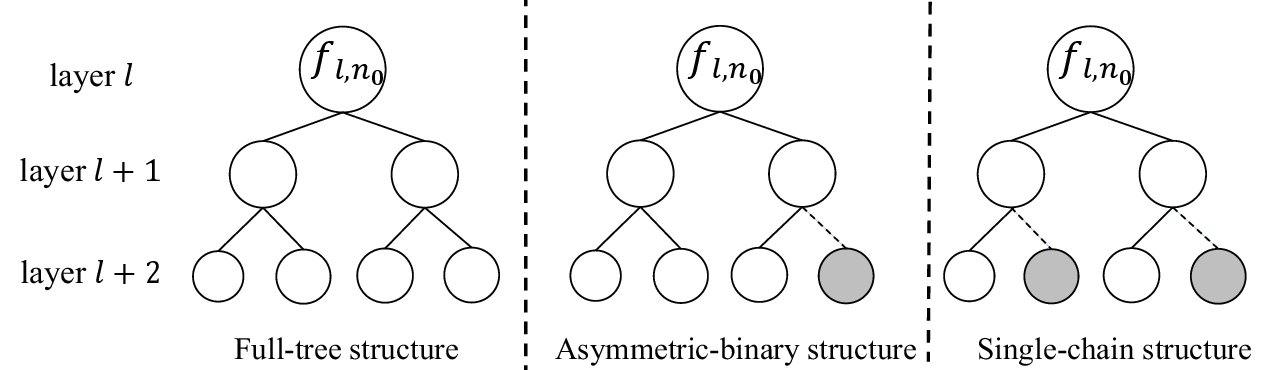}
	\caption{Three categories for topological structure of two-layer lookahead tree.}
	\label{fig5_2layer_lookahead}
\end{figure}

We firstly define the initial root node of search tree as the omnidirectional transmission, which is aligned to the two top-layer codewords dominant to beamspace $[-1,0]$ and $[0,1]$. The beginning search iteration counter is set as $t=-1$ and corresponding search layer is initialized to $l_{-1}=0$.
When BS forms directional beams in the $l_t$-th layer and determines the temporary index $n_0$ via UE feedback, the root node is updated and fixed as codeword $\bm f_{l_t,n_0}^+$. Then the strategy of the next $(t+1)$-th iteration can be categorized into three distinct cases based on the topological structure of tree as follow:
\begin{itemize}
	\item Full-tree structure: As shown in left of Fig. \ref{fig5_2layer_lookahead}, the two-step lookahead strategy aligns to a full binary tree. From the local perspective, there exist four leaf nodes to search, where the same number of timeslots are required whether hierarchical searching or bottom-level traversal is exploited ($T=4$). Therefore, in this case we directly adopt hierarchical training for consistency, i.e., $l_{t+1}=l_{t}+1$.
	
	\item Asymmetric-binary structure: As illustrated in the center of Fig. \ref{fig5_2layer_lookahead}, the left child node of the root is aligned to two leaf nodes denoted by $\bm f_{l_t+2,1}$ and $\bm f_{l_t+2,2}$, while the right child node is connected to single leaf node denoted by $\bm f_{l_t+2,3}$, where the codeword weight $\omega_{l_t,n}$ is derived by \eqref{weighted_codeword}. The exhaustive search at bottom layer will cost $T=3$ timeslots, whose reward is formulated as 
	\begin{equation}
		T_\text{ES} = 3(\omega_{l_t+2,1}+\omega_{l_t+2,2}+\omega_{l_t+2,3}).
	\end{equation}
	Besides, when we adopt hierarchical searching,  $T=4$ timeslots are required for reaching $\bm f_{l_t+2,1}$ and $\bm f_{l_t+2,2}$, while it cost $T=2$ timeslots for $\bm f_{l_t+2,3}$. Thus the expected overhead of hierarchical searching can be calculated as
	\begin{equation}
		T_\text{HS}=4\omega_{l_t+2,1}+4\omega_{l_t+2,2}+2\omega_{l_t+2,3}.
	\end{equation}
	And thus the most appropriate layer to be searched can be determined as
	\begin{equation}
		l_{t+1}=\left\{
		\arraycolsep=1.0pt\def\arraystretch{1.5}
		\begin{array}{cll}
		l_t + 1, && T_\text{HS}\leq T_\text{ES}\\
		l_t + 2, && T_\text{HS} > T_\text{ES}
		\end{array}
		\right. .
		\label{next_layer_piecewise}
	\end{equation} 
	\item Single-chain structure: As shown in the right of Fig. \ref{fig5_2layer_lookahead}, there are only two independent leaf nodes, allowing us to exhaustively transmit the two bottom-layer codewords to identify the dominant one, which enables the completion of the local search within the two-level lookahead. In this way, we have $l_{t+1} = l_{t}+2$.
\end{itemize}
In this way, the bottom-layer narrow beam can be ultimately determined with quite low complexity. The detailed expression is provided in Alg.\ref{low_comp_algo}.

\begin{algorithm}[tb] 
	\normalem
	\caption{Low-complexity 2-layer lookahead beam training} 
	\label{low_comp_algo} 
	\begin{algorithmic}[1] 
		\REQUIRE Training codebook $\{\bm f_{l,n}\}$, CKM $\mathcal{F}(\mathcal{P}|\bm f)$, UE positioning prior info $\bm{\mathcal{R}}$ and $\{P_{s}\}$, searching round $t=-1$.
		\ENSURE near-optimal beamforming codeword $\bm f_{L,n_\text{opt}}$
		
		\STATE Calculate beam potential $\{\omega_{l,n}\}$ via \eqref{weighted_codeword} similar to the initialization stage in Alg. \ref{single_user_algo}.
		\WHILE{currently searched layer $l_t<L$}
		\STATE Extract the 2-layer lookahead subtree and categorize its topological structure.
		\IF{full-tree structure}
		\STATE $l_{t+1}=l_{t}+1$.
		\ELSIF{Asymmetric-binary structure}
		\STATE Update the next layer $l_{t+1}$ via \eqref{next_layer_piecewise}.
		\ELSE
		\STATE $l_{t+1}=l_{t}+2$.
		\ENDIF
		
		\STATE BS searches the directional beams $\{\bm f_{l_{t+1},n}^+\}$.
		\STATE UE reports codeword index $n_{t+1}$ with largest power.
		\STATE Modify $\bm f_{l_{t+1},n_{t+1}}^+$ as the root node and update search tree as its child nodes.
		\STATE Searching time $t=t+1$.
		\ENDWHILE
	\end{algorithmic}
\end{algorithm}

\subsection{Performance analysis}

Given the spatial prior knowledge provided by CKM, the candidate beamspace to search can be drastically reduced. As a result, Alg.\ref{single_user_algo} outperforms traditional hierarchical beam training \cite{beam_training_hierarchical_4,beam_training_hierarchical_5} with lower overhead. In fact, the training overhead is difficult to quantify explicitly since it is positively correlated with potential beam distribution, which largely depends on the diversified UE position uncertainty. Since Alg.\ref{low_comp_algo} focuses on the local search tree structure rather than global reward, the overhead is slightly higher than that of Alg.\ref{single_user_algo}, yet still maintains an advantage over traditional methods \cite{beam_training_hierarchical_4,beam_training_hierarchical_5}.

As for computational complexity, we define $N_\text{beam}$ as the cardinality of initial potential beam set $\{\bm f_{L,n}^+\}$ generated by CKM. We neglect the CKM-aided initialization stage since it can be precalculated before beam training. For Alg.\ref{single_user_algo}, the detailed complexity mainly lies in the multiplications inside Stage 2. Overall $|\mathcal{A}|\lesssim N_\text{BS}$ activation cases and all $N_\text{beam}$ potential beams $\{\bm f_{L,n}^+\}$ should be separately considered to calculate individual overhead $\Xi_{z,n}$ in \eqref{overhead_calc} during almost $L$ rounds. Thus the overall computational complexity of Alg.\ref{single_user_algo} is yielded to $\mathcal{O}[N_\text{BS}N_\text{beam}^2(\log_2N_\text{BS})^2]$. Since Alg.\ref{low_comp_algo} focuses on the two-layer lookahead tree structure within almost $L$ rounds, the complexity can be derived as $\mathcal{O}[\log_2N_\text{BS}]$.

\section{Multi-user beam training extension}
Typical hierarchical beam training is confined to single-user scenario in practice. As for multi-user case, all users were independently searched, leading to  prohibitive overhead strictly proportional to the number of users, which constituted the primary bottleneck about multi-user training. In this section, we provide a development for multi-user joint hierarchical training scheme. By leveraging the codebook-oriented CKM, we establish a unified search framework but also exploit inter-user interference and correlation characteristics to realize collaborative improvement with reduced overhead. The key mechanism is that, during beam search, other users can gather auxiliary side information via current beam sidelobes from NLoS paths, where harnessing this information boosts the overall data volume and speeds up searching efficiency. In practice, by capturing sidelobe signals/energy from NLoS paths and processing them along with prior information from CKM, we can eliminate position grid points with weak beam gain correlation, enabling the rapid extraction and determination of users' positions and supporting beams. We implement the subscript $k$ to denote the user index and the detailed approach is outlined below.

\subsection{Multi-user training design}

\subsubsection{Searching layer determination}
\

In the initialization stage, similarly to the single-user scenario, we establish $K$ binary trees for overall users and calculate the corresponding codeword weights $\{\omega_{k,l,n}\}$ via \eqref{bottom_layer_weight} and \eqref{weighted_codeword}. The activation styles $\bm \alpha_z$ and $\mathcal{L}_z$ are generated the same as \eqref{activation_style}. Then the corresponding rewards can be further determined as $\mathcal{R}_{k,z,0}$ via \eqref{reward_1}, and thus the initial jointly-searching layer for overall users can be reformulated to
\begin{equation}
	l_0^\text{MU} = \min\left\{ \argmax_{\mathcal{L}_z} \sum_{k=1}^K \frac{\mathcal{R}_{k,z,0}}{\sum_{z=1}^{|\mathcal{A}|}\mathcal{R}_{k,z,0}} \right\},
	\label{MU_l}
\end{equation} 
where we adopt $l1$ normalization to eliminate the impact of user-specific large-scale fading on weights and rewards. Also, we calculate the single-user beginning layer $\{l_{0,k}^\text{SU}\}$ via \eqref{opt_layer}. Following the greedy principle, we select the highest layer to search, which is formulated as
\begin{equation}
	l_0^\text{opt} = \min \{l_0^\text{MU}, l_{0,1}^\text{SU},l_{0,2}^\text{SU},\dots,l_{0,K}^\text{SU}\}.
	\label{MU_opt_layer}
\end{equation}

When the beginning searching layer $l_0^\text{opt}$ is obtained, we need to determine the beams to search in current round. Comparing the values between $l_0^\text{opt}$ and user-specific layer $\{l_{0,k}^\text{SU} \}$, we can extract the matching UEs and divide them into three categories as follows:
\begin{equation}
\mathcal{I}_k = \left\{
	\arraycolsep=1.0pt\def\arraystretch{1.5}
	\begin{array}{cll}
		-1,&& l_{0,k}^\text{SU} > L\\
		0, &&  l_0^\text{opt}<l_{0,k}^\text{SU}\leq L\\
		1, && l_{0,k}^\text{SU} = l_0^\text{opt}
	\end{array}
\right. ,
\label{user_indicator}
\end{equation}
where $\mathcal{I}_k=1$ indicates user $k$ matches the current search layer, requiring its potential supportive directions to be considered in the beamspace searching. Conversely, for $\mathcal{I}_k=0$, the searching beams might not dominate in its supportive directions, which means only its gain distribution characteristics from beam sidelobes need attention. Besides, $\mathcal{I}_k=-1$ denotes the bottom-layer beam has been obtained with search termination for current user $k$.

Assume overall possible beams for each user inside the $l_t^\text{opt}$-th layer as $\mathcal{F}_{k,t}=\{\bm f_{k,l_t^\text{opt},n}^+\}$. Then in the first round of multi-user joint searching, BS takes the union set $\mathcal{F}_t$ of those matching users and forms these directional beams and probe them in several consecutive timeslots sequentially as:
\begin{equation}
\mathcal{F}_t = \bigcup_{ \mathcal{I}_k=1} \mathcal{F}_{k,t},\ \ |\mathcal{F}_t|\geq 2.
\label{searched_beam}
\end{equation}

After users receive the directional signals, the received signal powers corresponding to sequentially transmitted beam codewords are fed back to the BS, which is defined mathematically as $\bm g^\text{obs}_{k,t}\in \mathbb{R}^{|\mathcal{F}_t|\times 1}$ from \eqref{received_signal}:
\begin{equation}
\bm g^\text{obs}_{k,t} = \big(\ |y_k(\bm f_1)|,|y_k(\bm f_2)|,\dots,|y_k(\bm f_{|\mathcal{F}_t|})|\ \big)^T, \bm f_n\in \mathcal{F}_t.
\end{equation}
Correspondingly, the beam with strongest power can be formulated as 
\begin{equation}
\bm f^\text{obs}_{k,t} = \argmax_{\bm f_n\in \mathcal{F}_t} \ \big|y_k(\bm f_n)\big|,\  \mathcal{I}_k=1.
\end{equation}

\begin{figure}[!t]
	\centering
	\includegraphics[width=1\linewidth]{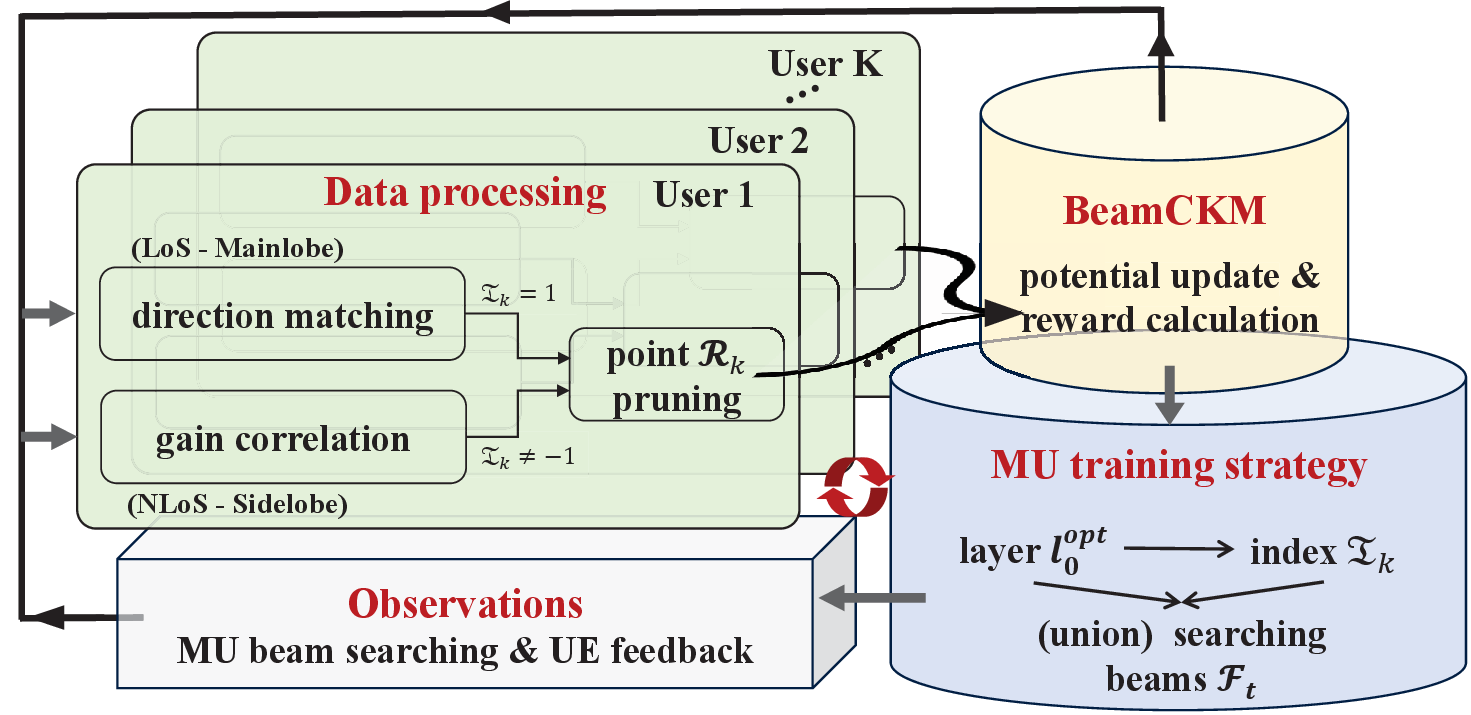}
	\caption{Block diagram of multi-user joint correlation-motivated position-pruning hierarchical training scheme.}
	\label{flow_MU_training}
\end{figure}

\subsubsection{Correlation-based position point pruning}
\

To facilitate distinct operations for the two categories users with $\mathcal{I}_k=1$ and $\mathcal{I}_k=0$, we need to design adaptive strategies in the pro-processing at BS side. Before that we firstly extract some useful information from CKM. Given the potential grid points $\bm p_{k,s,m}$ in \eqref{contiguous_region} and searched beams $\mathcal{F}_t$ in \eqref{searched_beam}, we can correspondingly calculate the pre-recorded CKM gain at each potential point $\bm p_{k,s,m}$ via \eqref{gain_map_CKM} as $\bm g^\text{map}_{k,t,s,m}\in\mathbb{R}^{|\mathcal{F}_t|\times 1}$:
\begin{equation}
	\bm g^\text{map}_{k,t,s,m} = \bigg[\ G_\text{map}(\bm p_{k,s,m},\bm f_1),\dots,G_\text{map}(\bm p_{k,s,m},\bm f_{|\mathcal{F}_t|})\ \bigg]^T,
	\label{MAP_gain}
\end{equation}
and the optimal beam inside set $\mathcal{F}_t$ for user $k$ at point $(s,m)$ is calculated as:
\begin{equation}
	\bm f^\text{map}_{k,t,s,m} = \argmax_{\bm f_n\in \mathcal{F}_t} \ G_\text{map}(\bm p_{k,s,m},\bm f_n), \ \mathcal{I}_k=1.
	\label{MAP_beam}
\end{equation}

As for users $\mathcal{I}_k=0$, we only focus on the observed gain distribution $\bm g^\text{obs}_{k,t}$ rather than the specific direction of the strongest beam, since the search beam set in such case might not include the optimal codeword for the current user. With real-time accurate observations as the reference, we can continuously eliminate mismatched position points inside CKM, until the unique beam direction is determined. Firstly, we adopt the cosine similarity as the correlation metric between observations and CKM data, where the detailed expression is 
\begin{equation}
	\Upsilon_{k,t,s,m} = \frac{\bm g^\text{obs \ H}_{k,t}\cdot \bm g^\text{map}_{k,t,s,m}}{|\bm g^\text{obs}_{k,t}|\cdot|\bm g^\text{map}_{k,t,s,m}|}.
	\label{correlation_calc}
\end{equation}
We dynamically filter out similarity values below a proportion-based cutoff threshold relative to the maximum value (e.g., $\eta=0.3$ times the maximum) to reduce feasible position points, where the pruned potential point set for users $\mathcal{I}_k=0$ can be formulated as
\begin{equation}
	\mathcal{R}_{k,t+1} = \big\{\bm p\in  \mathcal{R}_{k,t}| \Upsilon_{k,t,s,m}>\eta\times \max(\Upsilon_{k,t,s,m})\big\}.
	\label{Potential_point_set_Upsilon_0}
\end{equation}

As for users $\mathcal{I}_k=1$, both the observed gain distribution and the direction of dominant beam constitute essential information for training. Besides the aforementioned cosine similarity-based gain correlation evaluation, the dominant direction must be strictly consistent to the observation as for training accuracy. Therefore, the point-pruning strategy for users $\mathcal{I}_k=1$ should be rewritten as 
\begin{equation}
		\mathcal{R}_{k,t+1} = \bigg\{\bm p\in  \mathcal{R}_{k,t}\bigg|
			\arraycolsep=1.0pt\def\arraystretch{1.5}
		\begin{array}{ccc}
			&& \Upsilon_{k,t,s,m}>\eta\times \max(\Upsilon_{k,t,s,m}),\\
		&& \ \bm f_{k,t}^\text{obs}=\bm f_{k,t,s,m}^\text{map}
	\end{array}
\bigg\}.
\label{Potential_point_set_Upsilon_1}
\end{equation}

\subsubsection{Beam-tree update and iteration}
\

Following the above position-pruning strategies based on beam searching observations, we can update the potential position points for overall users $k=1,\dots,K$. Thus the beam search trees can be correspondingly adjusted also with their codeword weights similarly to stage 1 of Alg.\ref{single_user_algo}, which completes one round of multi-user beam training procedure. 

The overall searching continues until that overall users contain indicators $\mathcal{I}_k=-1, k=1,\dots,K$, i.e., all users are searched with optimal narrow bottom-layer codewords. The detailed expression is provided in Alg.\ref{multi_user_algo}.

\begin{algorithm}[htb] 
	\normalem
	\caption{Multi-user CKM-aided beam training} 
	\label{multi_user_algo} 
	\begin{algorithmic}[1] 
		\REQUIRE Codebook $\{\bm f_{l,n}\}$, CKM $\mathcal{F}(\mathcal{P}|\bm f)$, UE positioning prior info $\bm{\mathcal{R}}_k$ and $\{P_{k,s}\},k=1,\dots,K$, searching round $t=0$, hierarchical layer $L$, indicator $\mathcal{I}_k=1$. 
		\ENSURE optimal beamforming codeword $\bm f_{k,L,n_\text{opt}}$ for all users
		
		\WHILE{$\exists\  \mathcal{I}_k\neq -1,\ \ k=1,\dots,K $}
		\leftline{\ \ \ \ \   \textbf{\% Stage 1: layer determination and searching}}
		\STATE Calculate $\{l_{t,k}^\text{SU}\}$ and  $l_t^\text{opt}$ via \eqref{opt_layer} and \eqref{MU_opt_layer}. 
		\STATE Derive $\mathcal{I}_k$ via \eqref{user_indicator}.
		
		\STATE Get union beam set $\mathcal{F}_t$ of users  $\mathcal{I}_k=1$ in layer $l_t^\text{opt}$.
		
		\STATE BS searches the direction beams and UEs send received observation $\bm g_{k,t}^\text{obs}$ and $\bm f_{k,t}^\text{obs}$ back.
		
		\leftline{\ \ \ \ \   \textbf{\% Stage 2: Correlation-based point pruning}}
		
		\FOR{all users $k=1,\dots,K$}
		
		\STATE Derive $\bm g_{k,t,s,m}^\text{map}$, $\bm f^\text{map}_{k,t,s,m}$ from CKM via \eqref{MAP_gain}, \eqref{MAP_beam}.
		
		\STATE Calculate cosine similarity $\Upsilon_{k,t,s,m}$ between $\bm g_{k,t}^\text{obs}$ and $\bm g_{k,t,s,m}^\text{map}$ via \eqref{correlation_calc}.
		
		\STATE Update point set $\mathcal{R}_{k,t+1}$ via \eqref{Potential_point_set_Upsilon_0} or \eqref{Potential_point_set_Upsilon_1}.
		
		\ENDFOR

		\leftline{\ \ \ \ \   \textbf{\% Stage 3: beam tree update and iteration}}
		\STATE Update search trees and codeword weights for all users.
		\STATE searching time $t=t+1$.
		\ENDWHILE

	\end{algorithmic}
\end{algorithm}

\subsection{Performance analysis}

Alg.\ref{multi_user_algo} is implemented along the general framework of Alg.\ref{single_user_algo}. 
In contrast to Alg.\ref{single_user_algo}, which is restricted to the beam level, Alg.\ref{multi_user_algo} extends to the more fundamental level about UE spatial positions. 
The sidelobe gains from inter-user interference are additionally exploited, which is integrated with CKM, to support sufficient side information during beam training. And then all users' feasible position, also with optimal beam directions, can be simultaneously searched. In this way, Alg.\ref{multi_user_algo} outperforms Alg.\ref{single_user_algo} and other conventional schemes from the perspective of training overhead.

The detailed complexity for Alg.\ref{multi_user_algo} can be discussed separately across three stages. First, the initial layer determination contains complexity $\mathcal{O}[KN_\text{BS}N_\text{beam}^2(\log_2 N_\text{BS})^2]$ similarly to Alg.\ref{single_user_algo}. Then in Stage 2, the cosine similarity is calculated for all potential grid points in each searching round, which contains multiplication order $\mathcal{O}[N_\text{beam}\log_2 N_\text{BS}\cdot\sum_{k,s}N_{k,s}]$. In Stage 3, we need to update the beam tree during each searching round, where the codewords and corresponding weights should be updated to eliminate the influence of irrelevant grid positions from CKM. The complexity can be derived as $\mathcal{O}[N_\text{BS}\log_2 N_\text{BS}\cdot\sum_{k,s}N_{k,s}]$. Since we have $N_\text{beam}\ll N_\text{BS}$, the overall computational complexity is derived as $\mathcal{O}[KN_\text{BS}N_\text{beam}^2(\log_2 N_\text{BS})^2+N_\text{BS}\log_2 N_\text{BS}\cdot\sum_{k,s}N_{k,s}]$, which demands greater computational resources and complexity than Alg.\ref{single_user_algo}.

\section{Simulation Results}

\subsection{Scenario and Parameter Setting}
\begin{figure}[!t]
	\centering
	\includegraphics[width=0.7\linewidth]{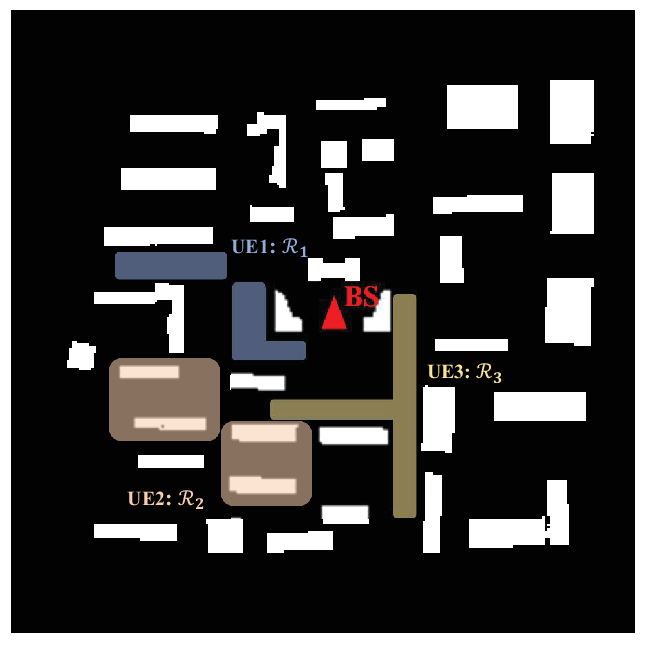}
	\caption{An illustration of the environment layout and transceiver range.}
	\label{fig6_simulation_setting}
\end{figure}

In this section, we investigate the performance of the proposed CKM-aided hierarchical beam training algorithms, in terms of the training overhead and corresponding spectral efficiency. The detailed illustration of the environment layout and transceiver range is shown in Fig. \ref{fig6_simulation_setting}, where the whole region $X\times Y=256\text{m}\times 256\text{m}$ is uniformly quantized into multiple grid points along the horizontal and vertical directions with $\Delta x=\Delta y=1\text{m}$. The multi-antenna BS is deployed at the center, with $128$-element ULA configuration and carrier frequency $80\text{GHz}$. $K=3$ single-antenna users are located in their respective regions, which are known in advance as prior information, derived from historical and sensing data. The potential region corresponding to individual user in Fig. \ref{fig6_simulation_setting} is rendered with different colors, and all users are supposed with random distribution inside those regions. As for the propagation channel, we assume the specular reflection and refraction exist while diffuse reflection does not, where all CKMs corresponding to different beam codewords are accurately calculated via Sionna ray-tracing. The geospatial representations undergo electromagnetic property set in Blender through randomized material parameterization before integration into the Sionna environment. As for the searching procedure, the total hierarchical layer is calculated as $L=7$ and the number of codewords inside each layer is $2^l,l=1,\dots,7$. The correlation threshold of \eqref{Potential_point_set_Upsilon_0} is set as $\eta=0.9$ for Alg. \ref{multi_user_algo}. The beam training process cycle is repeated 10,000 times, and average metric values are calculated to ensure the reliability.

\subsection{Single-user Training}

\begin{figure}[!t]
	\centering
	\includegraphics[width=1\linewidth]{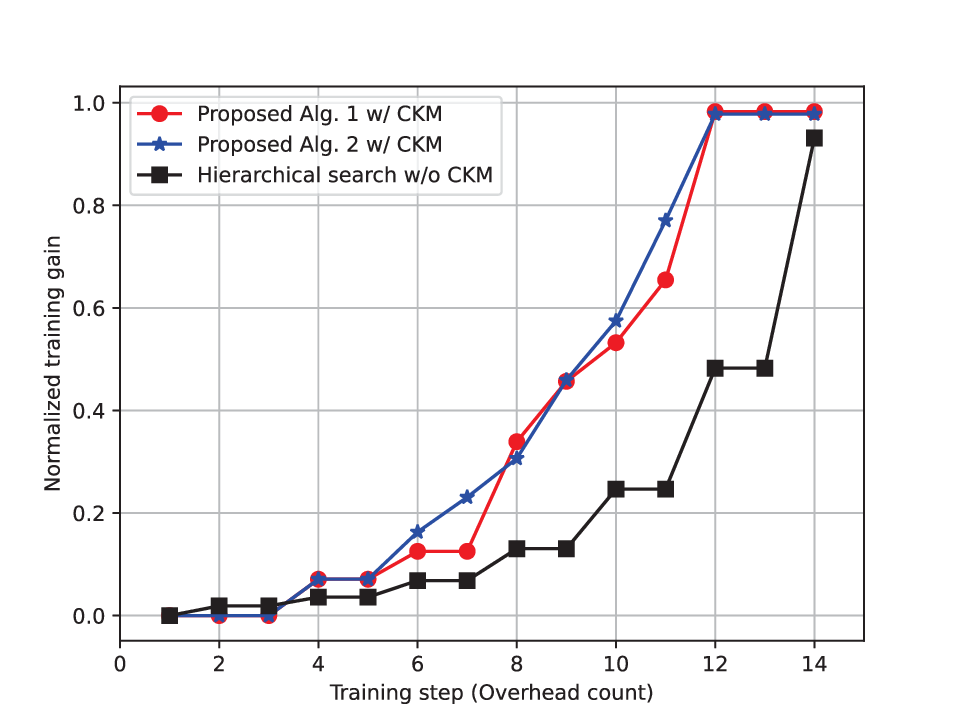}
	\caption{Comparison of average training during each search step for single-user scenario with $\text{SNR}=10\text{dB}$.}
	\label{fig7_SU_Gain_Overhead}
\end{figure}

Fig. \ref{fig7_SU_Gain_Overhead} presents the training gain of proposed single-user training algorithms against training steps (as overhead count increase). The baseline hierarchical search without CKM shows a significant drop in gain, where $L=7$ layers should be searched sequentially via binary method, i.e., $14$ codewords. Compared with the conventional benchmark, the proposed Alg. \ref{single_user_algo} and Alg. \ref{low_comp_algo} require less training overhead and stronger training gain when search completed. This can be explained by the fact that CKM can provide sufficient prior information, which helps narrow down the search range and thereby reduces the required pilot overhead. Meanwhile, such prior information can avoid repeated bisection and update, thus reducing misjudgments with noise and improving the final training gain.

\begin{figure}[!t]
	\centering
	\includegraphics[width=1\linewidth]{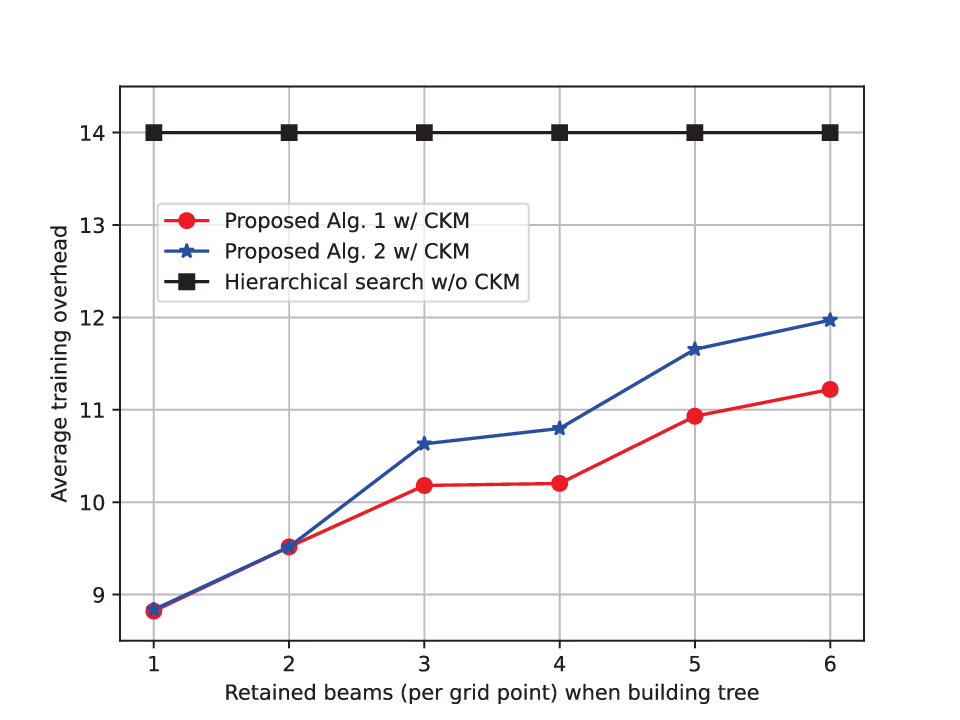}
	\caption{Comparison of overhead consumption against retained beams during building tree for single-user scenario with $\text{SNR}=10\text{dB}$.}
	\label{fig8_SU_Overhead_NumBeam}
\end{figure}

In fact it is difficult to evaluate the impact of potential UE ranges on the beam training overhead since the impact of regions may be affected by several factors such as shape, area, and corresponding CKM beam distribution, all of which exert complex influences on the results and thus it cannot be quantitatively characterized. To address this, we fix the potential range of UE and instead retain different numbers of beams when constructing the search tree in \eqref{bottom_layer_weight}, thereby evaluating the adaptability and efficiency of the proposed schemes. Fig. \ref{fig8_SU_Overhead_NumBeam} explores the relationship between training overhead and the number of retained beams per grid points 1$\sim$6 when building tree in \eqref{bottom_layer_weight}. As shown here, the baseline hierarchical search without CKM exhibits a stable overhead line, as it always requires exhaustive searching of all candidate beams. In contrast, the CKM-assisted algorithms  maintain lower overhead consumption across all retained beam numbers. This advantage arises from CKM’s ability to pre-select optimal beam candidates from the codebook, avoiding the requirement to probe all directions. Besides, we can find that when only retaining single beam for each grid point, Alg.\ref{single_user_algo} and Alg.\ref{low_comp_algo} contain almost the same overhead which is consistent to Fig. \ref{fig7_SU_Gain_Overhead}. As the number of retained beams grows,  a difference emerges in the training overhead required by these two algorithms that the low-complexity Alg.\ref{low_comp_algo} relies on higher training overhead, while it still maintains advantage over the traditional hierarchical search without CKM.

\begin{figure}[!t]
	\centering
	\includegraphics[width=1\linewidth]{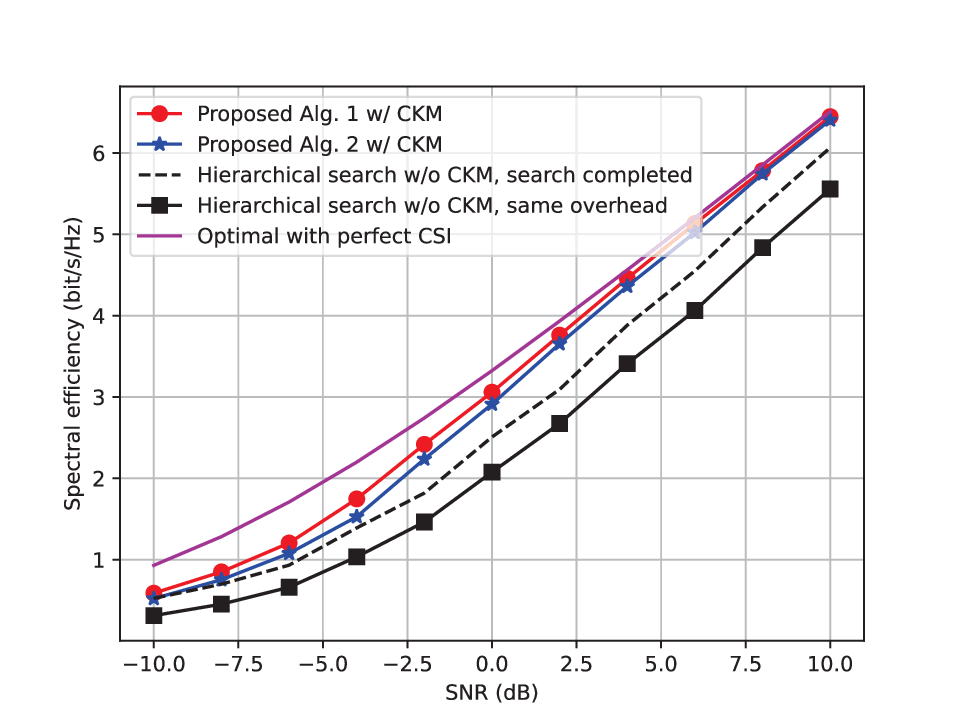}
	\caption{Comparison of spectral efficiency against SNR under different training strategies for single-user scenario.}
	\label{fig9_SU_SE_SNR}
\end{figure}

Fig. \ref{fig9_SU_SE_SNR} compares the spectral efficiency under varying SNR. The optimal SE with pre-determined CSI and the strongest beam codeword is set as benchmark and upper bound. We can easily observe that, at all SNR levels, the CKM-assisted algorithms achieve higher SE values close to the optimal perfect CSI curve. Compared with the hierarchical training with full complete training, the spectral efficiency can be significantly improved which proves that its beam training accuracy and beam gain are enhanced. In practice, when we artificially set all methods to consume the same number of training overhead, the proposed method can further improve the spectral efficiency by approximately $30\%$. This also indirectly indicates that the proposed method has advantages in training overhead besides beam gain.

\subsection{Multi-user Scenario}

\begin{figure}[!t]
	\centering
	\includegraphics[width=0.9\linewidth]{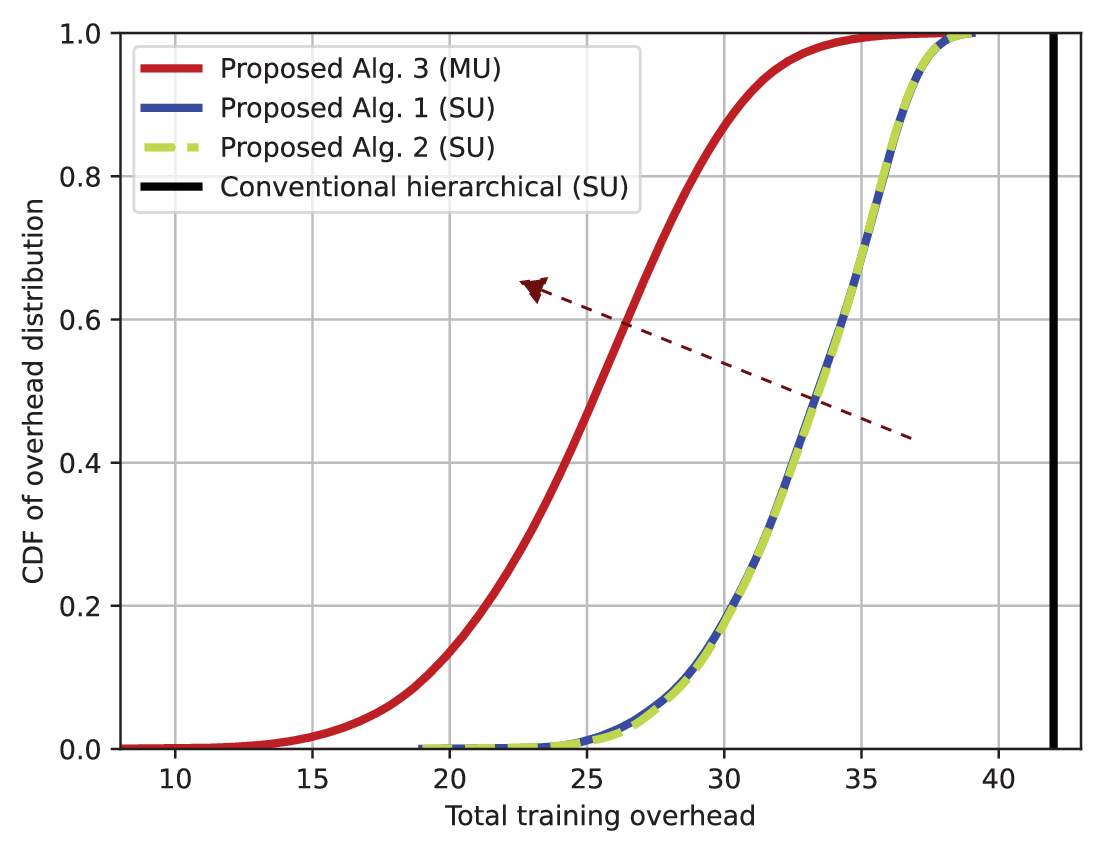}
	\caption{Comparison of CDF distributions for training overhead under different strategies for multi-user scenario with $\text{SNR}=5\text{dB}$.}
	\label{fig10_MU_CDF_overhead}
\end{figure}

\begin{figure}[!t]
	\centering
	\includegraphics[width=0.9\linewidth]{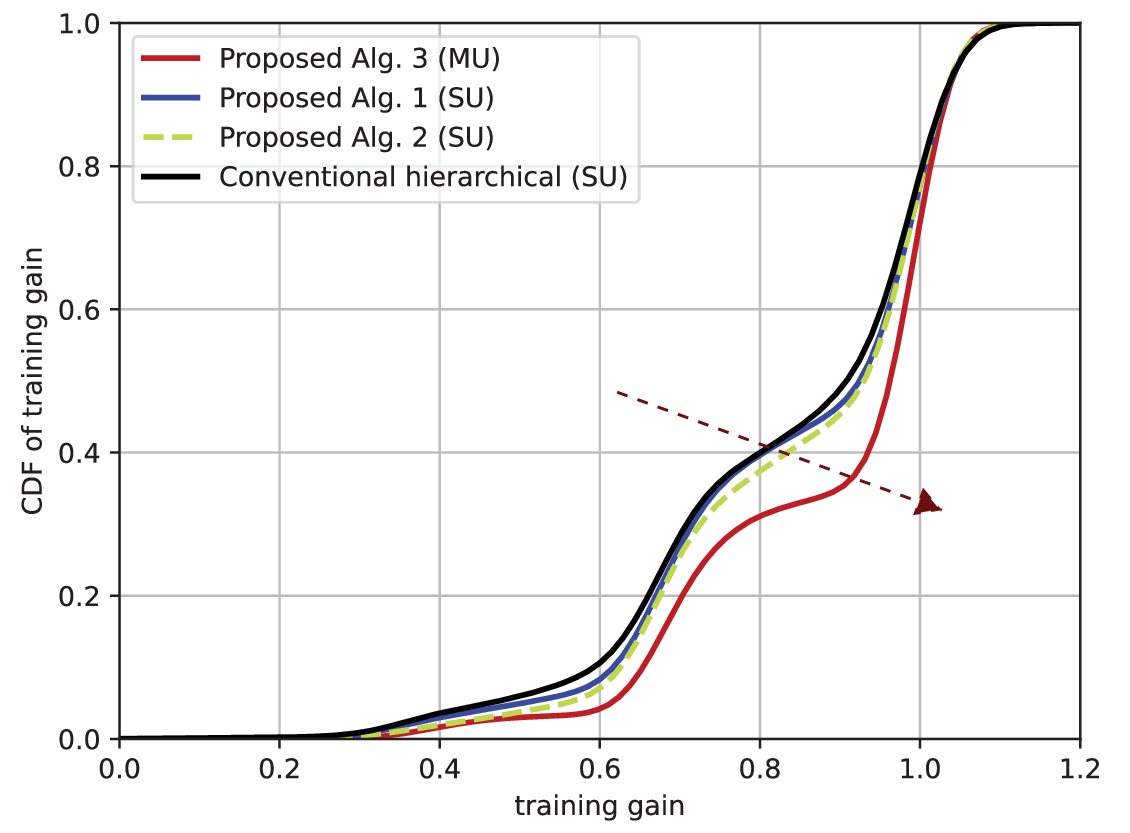}
	\caption{Comparison of CDF distributions for training gain under different strategies for multi-user scenario with $\text{SNR}=5\text{dB}$.}
	\label{fig11_MU_CDF_gain}
\end{figure}

As for the multi-user scenario, we only focus on the training performance such as the total overhead and training gain. This implies that subsequent beamforming optimization is no longer within the scope of the present study. The reason lies in the fact that in multi-user scenarios, there are occasional instances where the dominant beams of several users (randomly generated) coincide, and the further quantitative analysis of beam conflicts and inter-user interference is necessary. In contrast, the current work focuses exclusively on the beam training process for all users. Nevertheless, it is worth noting that the proposed schemes can offer a solution to this specific case. Specifically, each user can retain two strongest candidate beam codewords rather than one during the CKM-aided training process, thereby providing a solution for subsequent beam switch and alignment.

Fig. \ref{fig10_MU_CDF_overhead} presents the cumulative distribution function (CDF) of total training overhead, directly reflecting the efficiency of beam training across SU and MU algorithms.  As shown here, it translates to almost $20\%$ reduction in overhead for single-user methods like Alg.\ref{single_user_algo} and Alg.\ref{low_comp_algo} when CKM is integrated. Almost $60\%$ trials can achieve overhead levels within $T=35$ for three users. More notably, the multi-user jointly training scheme (Proposed Alg.\ref{multi_user_algo}) outperforms the two single-user algorithms by a significant gap. Nearly $100\%$ of trials for Proposed Alg.3 (MU) result in a total training overhead of $35$, and a large proportion of trials even achieve overhead levels comparable to the most efficient CKM-assisted SU algorithms. The reason for this MU advantage lies in CKM’s ability to enable inter-user beam coordination. Instead of performing independent beam training for each user, the inter-user interference and beam sidelobes are utilized to explore the implicit side information to assist search process. Specifically, by leveraging the side-lobe gain distribution of beams from NLoS paths, and effectively combining them with CKM via cosine similarity calculations, more feasible grids with higher correlation can be filtered out rapidly to reduce the search range and training overhead, at the cost of slightly complicated user feedback.

Fig. \ref{fig11_MU_CDF_gain} depicts the CDF of training gain along $10000$ trials. All simulation schemes conducted full beam training (i.e., without considering overhead issues) and the final beam distributions exhibit similar and comparable distribution patterns. Nevertheless, it is still evident that the proposed multi-user joint CKM-aided beam training method yields significantly stronger training gains. Meanwhile, when considered in conjunction with Fig. \ref{fig10_MU_CDF_overhead} above, the proposed Alg.\ref{multi_user_algo} demonstrates superior and more robust performance advantages in both training overhead and searching gains.

\section{Conclusions}
This paper addressed the limitations hindering the practical use of CKM-aided beam training in 6G systems, specifically solving the problems of unknown user positions before training and unclear interplay among CKM, observations, and training strategies. For the single-user scenario, we proposed a reward-motivated beam-potential hierarchical strategy that integrated partial position information with CKM via potential codeword weights and rewards. And it was complemented by a low-complexity 2-layer lookahead scheme to balance overhead and computational demand. In the multi-user scenario, we developed a correlation-driven position-pruning training scheme that exploited sidelobe gains from inter-user interference as side information, reducing training overhead while enabling simultaneous assignment of optimal beams to all users. Simulation results confirmed the superiority of the proposed approaches. We tried to solidify the practical value of CKM-aided beam training for 6G, and future work might extend the frameworks to dynamic user mobility or ultra-dense networks, as well as explore integrating reinforcement learning for real-time adaptive optimization of training strategies.


%

%
%
%
%
%

\bibliographystyle{ieeetr}
\normalem
\bibliography{bare_jrnl_v1.bib}

@article{survey1,
  title={{Towards 6G wireless communication networks: Vision, enabling technologies, and new paradigm shifts}},
  author={You, Xiaohu and Wang, Cheng-Xiang and Huang, Jie and Gao, Xiqi and Zhang, Zaichen and Wang, Mao and Huang, Yongming and Zhang, Chuan and Jiang, Yanxiang and Wang, Jiaheng and others},
  journal={Sci. China Inf. Sci.},
  volume={64},
  number={1},
  pages={110301},
  year={Nov. 2021},
  publisher={Springer}
}

@article{survey2,
  title={{Vision, requirements, and technology trend of 6G: How to tackle the challenges of system coverage, capacity, user data-rate and movement speed}},
  author={Chen, Shanzhi and Liang, Ying-Chang and Sun, Shaohui and Kang, Shaoli and Cheng, Wenchi and Peng, Mugen},
  journal={IEEE Wirel. Commun.},
  volume={27},
  number={2},
  pages={218--228},
  year={April 2020},
  publisher={IEEE}
}

@article{survey3,
  title={{A tutorial on environment-aware communications via channel knowledge map for 6G}},
  author={Zeng, Yong and Chen, Junting and Xu, Jie and Wu, Di and Xu, Xiaoli and Jin, Shi and Gao, Xiqi and Gesbert, David and Cui, Shuguang and Zhang, Rui},
  journal={IEEE Commun. Surv. Tutor.},
  volume={26},
  number={3},
  pages={1478--1519},
  year={Feb. 2024},
  publisher={IEEE}
}

@ARTICLE{survey4,
	author={Wang, Heng and Zhang, Jianhua and Nie, Gaofeng and Yu, Li and Yuan, Zhiqiang and Li, Tongjie and Wang, Jialin and Liu, Guangyi},
	journal={IEEE Commun. Mag.}, 
	title={{Digital Twin Channel for 6G: Concepts, Architectures and Potential Applications}}, 
	year={March 2025},
	volume={63},
	number={3},
	pages={24-30}
}

@article{survey5,
  title={{Channel Knowledge Maps for 6G Wireless Networks: Construction, Applications, and Future Challenges}},
  author={Liu, Xingchen and Sun, Shu and Tao, Meixia and Kaushik, Aryan and Yan, Hangsong},
  journal={arXiv preprint arXiv:2505.24151},
  year={2025}
}

@ARTICLE{CKM_constrct_0,
	author={Rizk, K. and Wagen, J.-F. and Gardiol, F.},
	journal={IEEE Trans. Veh. Technol.}, 
	title={{Two-dimensional ray-tracing modeling for propagation prediction in microcellular environments}}, 
	year={May 1997},
	volume={46},
	number={2},
	pages={508-518}
}

@article{CKM_constrct_1,
  title={{RadioUNet: Fast radio map estimation with convolutional neural networks}},
  author={Levie, Ron and Yapar, {\c{C}}a{\u{g}}kan and Kutyniok, Gitta and Caire, Giuseppe},
  journal={IEEE Trans. Wirel. Commun.},
  volume={20},
  number={6},
  pages={4001--4015},
  year={June 2021},
  publisher={IEEE}
}

@article{CKM_constrct_2,
  title={{RME-GAN: A learning framework for radio map estimation based on conditional generative adversarial network}},
  author={Zhang, Songyang and Wijesinghe, Achintha and Ding, Zhi},
  journal={IEEE Internet Things J.},
  volume={10},
  number={20},
  pages={18016--18027},
  year={Oct. 2023},
  publisher={IEEE}
}

@article{CKM_constrct_3,
  title={{WiFi-Diffusion: Achieving Fine-Grained WiFi Radio Map Estimation with Ultra-Low Sampling Rate by Diffusion Models}},
  author={Liu, Zhiyuan and Zhang, Shuhang and Liu, Qingyu and Zhang, Hongliang and Song, Lingyang},
  journal={IEEE J. Sel. Areas Commun.},
  year={July 2025},
  publisher={IEEE}
}

@inproceedings{CKM_constrct_4,
  title={{Channel knowledge map for environment-aware communications: EM algorithm for map construction}},
  author={Li, Kun and Li, Peiming and Zeng, Yong and Xu, Jie},
  booktitle={2022 IEEE Wireless Commun. Netw. Conf. (WCNC)},
  pages={1659--1664},
  year={2022},
  organization={IEEE}
}

@article{CKM_constrct_5,
  title={{RadioDiff-Inverse: Diffusion Enhanced Bayesian Inverse Estimation for ISAC Radio Map Construction}},
  author={Wang, Xiucheng and Fang, Zhongsheng and Cheng, Nan and Sun, Ruijin and Li, Zan and others},
  journal={arXiv preprint arXiv:2504.14298},
  year={2025}
}

@ARTICLE{CKM_constrct_6,
	author={Jin, Zhenzhou and You, Li and Wang, Jue and Xia, Xiang-Gen and Gao, Xiqi},
	journal={IEEE Trans. Wirel. Commun.}, 
	title={{An I2I Inpainting Approach for Efficient Channel Knowledge Map Construction}}, 
	year={Feb. 2025},
	volume={24},
	number={2},
	pages={1415-1429}
}

@article{CKM_constrct_ICI_7,
  title={{Interference-Cancellation-Based Channel Knowledge Map Construction and Its Applications to Channel Estimation}},
  author={Jiang, Wenjun and Yuan, Xiaojun and Teng, Boyu and Wang, Hao and Qian, Jing},
  journal={IEEE Trans. Wirel. Commun.},
  year={July 2025},
  volume={24},
  number={7},
  pages={6240-6256},
  publisher={IEEE}
}

@article{CKM_constrct_ICI_8,
  title={{IMNet: Interference-aware channel knowledge map construction and localization}},
  author={Zhao, Le and Fei, Zesong and Wang, Xinyi and Huang, Jingxuan and Li, Yuan and Zhang, Yan},
  journal={IEEE Wirel. Commun. Lett.},
  year={March 2025},
  volume={14},
  number={3},
  pages={856-860},
  publisher={IEEE}
}

@article{CKM_constrct_data_9,
  title={{How much data is needed for channel knowledge map construction?}},
  author={Xu, Xiaoli and Zeng, Yong},
  journal={IEEE Trans. Wirel. Commun.},
  volume={23},
  number={10},
  pages={13011--13021},
  year={Oct. 2024},
  publisher={IEEE}
}

@article{CKM_constrct_update_10,
  title={{Lightweight and Self-Evolving Channel Twinning: An Ensemble DMD-Assisted Approach}},
  author={Cao, Yashuai and Wang, Jintao and Shi, Xu and Ni, Wei},
  journal={IEEE Trans. Wirel. Commun.},
  year={Oct. 2025},
  volume={24},
  number={10},
  pages={8072-8085}
}

@article{CKM_constrct_MIMO_11,
  title={{Channel Fingerprint Construction for Massive MIMO: A Deep Conditional Generative Approach}},
  author={Jin, Zhenzhou and You, Li and Li, Xudong and Gao, Zhen and Liu, Yuanwei and Xia, Xiang-Gen and Gao, Xiqi},
  journal={arXiv preprint arXiv:2505.07893},
  year={2025}
}

@INPROCEEDINGS{CKM_constrct_MIMO_12,
AUTHOR="Haohan Wang and Xu Shi and Hengyu Zhang and Yashuai Cao and Jintao Wang",
TITLE={{Beamforming-Codebook-Aware Channel Knowledge Map Construction for
Multi-Antenna Systems}},
BOOKTITLE="IEEE Global Commun. Conf. (Globecom), Taipei, Taiwan",
MONTH=dec,
YEAR=2025,
}

@article{CKM_adopt_beamforming_1,
  title={{Environment-aware hybrid beamforming by leveraging channel knowledge map}},
  author={Wu, Di and Zeng, Yong and Jin, Shi and Zhang, Rui},
  journal={IEEE Trans. Wirel. Commun.},
  volume={23},
  number={5},
  pages={4990--5005},
  year={May 2024},
  publisher={IEEE}
}

@article{CKM_adopt_beamforming_2,
  title={{Channel Knowledge Map-assisted Dual-domain Tracking and Predictive Beamforming for High-Mobility Wireless Networks}},
  author={Du, Ruolin and Wei, Zhiqiang and Yang, Zai and Yang, Lei and Zeng, Yong and Ng, Derrick Wing Kwan and Yuan, Jinhong},
  journal={arXiv preprint arXiv:2506.22796},
  year={2025}
}

@article{CKM_adopt_beamforming_3,
	title={{Neural Channel Knowledge Map Assisted Scheduling Optimization of Active IRSs in Multi-User Systems}},
	author={Chen, Xintong and Jiang, Zhenyu and Lyu, Jiangbin and Fu, Liqun},
	journal={arXiv preprint arXiv:2508.07009},
	year={2025}
}

@ARTICLE{CKM_adopt_beamforming_4,
	author={Li, Bowen and Chen, Junting},
	journal={IEEE Trans. Wirel. Commun.}, 
	title={{Radio map assisted approach for interference-aware predictive UAV communications}},
	year={Nov. 2024},
	volume={23},
	number={11},
	pages={16725-16741}
}

@ARTICLE{CKM_adopt_CSI_5,
	author={Wu, Di and Qiu, Yuelong and Zeng, Yong and Wen, Fuxi},
	journal={IEEE Wirel. Commun. Lett.}, 
	title={{Environment-aware channel estimation via integrating channel knowledge map and dynamic sensing information}}, 
	year={Dec. 2024},
	volume={13},
	number={12},
	pages={3608-3612}
}

@ARTICLE{CKM_adopt_CSI_6,
	author={Wang, Xianling and Shi, Yi and Wang, Tianci and Huang, Yingyujiao and Hu, Zeyu and Chen, Lin and Jiang, Zhiyuan},
	journal={IEEE Trans. Commun.}, 
	title={{Channel Knowledge Map-Aided Channel Prediction With Measurements-Based Evaluation}}, 
	year={May 2025},
	volume={73},
	number={5},
	pages={3622-3636}
}

@ARTICLE{CKM_adopt_beam_8,
	author={Shao, Chenyang and Liu, Chunshan and Zhao, Lou and Li, Min and Zhang, Xiaoshuai and Sun, Minhong},
	journal={IEEE Wirel. Commun. Lett.}, 
	title={{Deep Learning-Based Millimeter Wave Beam Recommendation via Channel Knowledge Map}}, 
	year={June 2025},
	volume={14},
	number={6},
	pages={1648-1652}
}

@article{beam_training_1,
  title={{Higher spectral efficiency for mmWave MIMO: Enabling techniques and precoder designs}},
  author={Wang, Jintao and Zhang, Xiaohui and Shi, Xu and Song, Jian},
  journal={IEEE Commun. Mag.},
  volume={59},
  number={4},
  pages={116--122},
  year={April 2021},
  publisher={IEEE}
}

@ARTICLE{beam_training_2,
	author={Shi, Xu and Wang, Jintao and Wang, Xuehan and You, Changsheng and Song, Jian},
	journal={IEEE Trans. Commun.}, 
	title={{Double-Sided Near-Field XL-MIMO: Beamfocusing Codeword Selection and Channel Estimation}}, 
	year={May 2025},
	volume={73},
	number={5},
	pages={3441-3455}
}

@article{beam_training_exhaustive_3,
  title={{Beam codebook based beamforming protocol for multi-Gbps millimeter-wave WPAN systems}},
  author={Wang, Junyi and Lan, Zhou and Pyo, Chang-woo and Baykas, Tuncer and Sum, Chin-sean and Rahman, Mohammad Azizur and Gao, Jing and Funada, Ryuhei and Kojima, Fumihide and Harada, Hiroshi and others},
  journal={IEEE J. Sel. Areas Commun.},
  volume={27},
  number={8},
  pages={1390--1399},
  year={Oct. 2009},
  publisher={IEEE}
}

@article{beam_training_hierarchical_4,
  title={{Hierarchical codebook design for beamforming training in millimeter-wave communication}},
  author={Xiao, Zhenyu and He, Tong and Xia, Pengfei and Xia, Xiang-Gen},
  journal={IEEE Trans. Wirel. Commun.},
  volume={15},
  number={5},
  pages={3380--3392},
  year={May 2016},
  publisher={IEEE}
}

@article{beam_training_hierarchical_5,
  title={{Hierarchical codebook-based multiuser beam training for millimeter wave massive MIMO}},
  author={Qi, Chenhao and Chen, Kangjian and Dobre, Octavia A and Li, Geoffrey Ye},
  journal={IEEE Trans. Wirel. Commun.},
  volume={19},
  number={12},
  pages={8142--8152},
  year={Dec. 2020},
  publisher={IEEE}
}

@article{beam_training_hierarchical_6,
  title={{Wide-beam designs for terahertz massive MIMO: SCA-ATP and S-SARV}},
  author={Ning, Boyu and Wang, Tiantian and Huang, Chongwen and Zhang, Yuchen and Chen, Zhi},
  journal={IEEE Internet Things J.},
  volume={10},
  number={12},
  pages={10857--10869},
  year={June 2023},
  publisher={IEEE}
}

@article{beam_training_hierarchical_near_7,
  title={{Spatial-chirp codebook-based hierarchical beam training for extremely large-scale massive MIMO}},
  author={Shi, Xu and Wang, Jintao and Sun, Zhi and Song, Jian},
  journal={IEEE Trans. Wirel. Commun.},
  volume={23},
  number={4},
  pages={2824--2838},
  year={April 2024},
  publisher={IEEE}
}

@article{beam_training_multifinger_8,
  title={{Multi-user beam training and transmission design for covert millimeter-wave communication}},
  author={Zhang, Jiayu and Li, Min and Zhao, Min-Jian and Ji, Xiaoyu and Xu, Wenyuan},
  journal={IEEE Trans. Inf. Forensics Secur.},
  volume={17},
  pages={1528--1543},
  year={March 2022},
  publisher={IEEE}
}

@article{beam_training_multifinger_9,
  title={{Frequency-Scanning-Based Fast Multiuser Beam Training for Wideband Massive MIMO}},
  author={Shi, Xu and Wang, Jintao and Song, Jian},
  journal={IEEE Internet Things J.},
  volume={11},
  number={11},
  pages={19852--19865},
  year={June 2024},
  publisher={IEEE}
}

@article{beam_training_Kalman_10,
  title={{Fast channel tracking for terahertz beamspace massive MIMO systems}},
  author={Gao, Xinyu and Dai, Linglong and Zhang, Yuan and Xie, Tian and Dai, Xiaoming and Wang, Zhaocheng},
  journal={IEEE Trans. Veh. Technol.},
  volume={66},
  number={7},
  pages={5689--5696},
  year={July 2017},
  publisher={IEEE}
}

@article{beam_training_deeplearning_11,
  title={{Deep learning assisted mmWave beam prediction for heterogeneous networks: A dual-band fusion approach}},
  author={Ma, Ke and Du, Shouliang and Zou, Haoming and Tian, Wenqiang and Wang, Zhaocheng and Chen, Sheng},
  journal={IEEE Trans. Commun.},
  volume={71},
  number={1},
  pages={115--130},
  year={Jan. 2022},
  publisher={IEEE}
}

@article{beam_training_deeplearning_12,
  title={{Deep learning for mmWave beam-management: State-of-the-art, opportunities and challenges}},
  author={Ma, Ke and Wang, Zhaocheng and Tian, Wenqiang and Chen, Sheng and Hanzo, Lajos},
  journal={IEEE Wirel. Commun.},
  volume={30},
  number={4},
  pages={108--114},
  year={Aug. 2022},
  publisher={IEEE}
}

@ARTICLE{beam_training_side_info_13,
	author={Liu, Linyang and You, Changsheng and Zhang, Yunpu and Liu, Tianyu},
	journal={IEEE Commun. Lett.}, 
	title={Side Angle Information Assisted Near-Field Beam Training for XL-Array Communications}, 
	year={Sep. 2024},
	volume={28},
	number={9},
	pages={2201-2205}
}

@inproceedings{beam_training_side_info_14,
  title={{Side-information-aided noncoherent beam alignment design for millimeter wave systems}},
  author={Zhang, Yi and Patel, Kartik and Shakkottai, Sanjay and Jr, Robert W Heath},
  booktitle={Proc. 20th ACM Int. Symp. Mob. Ad Hoc Netw. Comput.},
  pages={341--350},
  year={July 2019}
}

@article{beam_training_environment_15,
  title={{Environment-specific beam training for extremely large-scale MIMO systems via contrastive learning}},
  author={Zhang, Xiangyu and Zhang, Haiyang and Li, Chunguo and Huang, Yongming and Yang, Luxi},
  journal={IEEE Commun. Lett.},
  volume={27},
  number={10},
  pages={2638--2642},
  year={Oct. 2023},
  publisher={IEEE}
}

@ARTICLE{GPS_error_1,
  author={Liu, Zhidan and Liu, Jiancong and Xu, Xiaowen and Wu, Kaishun},
  journal={IEEE Trans. Mob. Comput.}, 
  title={{DeepGPS: Deep Learning Enhanced GPS Positioning in Urban Canyons}}, 
  year={Jan. 2024},
  volume={23},
  number={1},
  pages={376-392},
}

@ARTICLE{WIFI_error_2,
  author={Xue, Weixing and Qiu, Weining and Hua, Xianghong and Yu, Kegen},
  journal={IEEE Sens. J.}, 
  title={{Improved Wi-Fi RSSI Measurement for Indoor Localization}}, 
  year={April 2017},
  volume={17},
  number={7},
  pages={2224-2230},
}

@ARTICLE{highspeed_error_3,
  author={Zhou, Tao and Zhang, Haitong and Ai, Bo and Xue, Chen and Liu, Liu},
  journal={IEEE Trans. Wirel. Commun.}, 
  title={{Deep-Learning-Based Spatial–Temporal Channel Prediction for Smart High-Speed Railway Communication Networks}}, 
  year={July 2022},
  volume={21},
  number={7},
  pages={5333-5345},
}

@ARTICLE{Disaster_error_4,
  author={Zhou, Yi and Jin, Zhanqi and Shi, Huaguang and Shi, Lei and Lu, Ning and Dong, Mianxiong},
  journal={IEEE Trans. Netw. Sci. Eng.}, 
  title={{Enhanced Emergency Communication Services for Post–Disaster Rescue: Multi-IRS Assisted Air-Ground Integrated Data Collection}}, 
  year={Oct. 2024},
  volume={11},
  number={5},
  pages={4651-4664},
}

\ifCLASSOPTIONcaptionsoff
\newpage
\fi

\end{document}